\documentclass[usenatbib,usegraphicx]{mn2e}
\usepackage{times}
\usepackage{graphicx, epsfig}
\usepackage[fleqn]{amsmath}
\usepackage{amsfonts}
\usepackage{amssymb}
\usepackage[totalwidth=17.8cm, totalheight=24.0cm]{geometry}

\title[The SAURON project XII]{The SAURON project - XII. Kinematic
  substructures in early-type galaxies: evidence for disks in fast
  rotators}

\author[Davor Krajnovi\'c et al.]  {Davor
  Krajnovi\'c,$^1$\thanks{E-mail: dxk@astro.ox.ac.uk} R.\ Bacon,$^2$
  Michele Cappellari,$^1$ Roger L.\ Davies,$^1$ \newauthor P.\ T.\ de
  Zeeuw,$^{3,9}$ Eric Emsellem,$^2$ Jes\'us Falc\'on-Barroso,$^{4}$
  \newauthor Harald Kuntschner,$^5$ Richard M. McDermid,$^6$ Reynier
  F.\ Peletier,$^{7}$ \newauthor Marc Sarzi,$^8$ Remco C. E. van den Bosch$^9$and  Glenn van de
  Ven$^{10}$\thanks{Hubble Fellow}  \\
$^1$Denys Wilkinson Building, University of Oxford, Keble Road, OX1 3RH, UK\\
$^2$Universit\'e de Lyon, France; Universit\'e Lyon 1, F-69007; CRAL, Observatoire
de Lyon, F-69230 Saint Genis Laval; CNRS, UMR 5574 ; ENS de Lyon, France \\
$^3$European Southern Observatory, Karl-Schwarzschild-Str~2, 85748 Garching, Germany\\
$^4$European Space and Technology Centre (ESTEC), Keplerlaan 1, Postbus 299, 2200 AG Noordwijk, The Netherlands \\
$^5$Space Telescope European Coordinating Facility, European Southern Observatory, Karl-Schwarzschild-Str~2, 85748 Garching, Germany\\
$^6$Gemini Observatory, Northern Operations Center, 670 N. A'ohoku Place,Hilo, Hawaii, 96720, USA\\
$^7$Kapteyn Astronomical Institute, Postbus 800, 9700 AV Groningen, The Netherlands \\
$^8$Centre for Astrophysics Research, University of Hertfordshire, Hatfield, Herts AL1 09AB\\
$^9$Sterrewacht Leiden, Leiden University, Niels Bohrweg~2, 2333~CA Leiden, The Netherlands\\
$^{10}$Institute for Advanced Study, Peyton Hall, Princeton, NJ 08544, USA}

\pagerange{\pageref{firstpage}--\pageref{lastpage}}
\pubyear{2007}

\newcommand{\kms}{\>{\rm km}\,{\rm s}^{-1}}

\newcommand{\PAk}{{\rm PA}$_{kin}$}
\newcommand{\qk}{{\rm q}$_{kin}$}
\newcommand{\qp}{{\rm q}$_{phot}$}
\newcommand{\PAp}{{\rm PA}$_{phot}$}
\newcommand{\kinemetry}{{\it kinemetry}}

\def\aj{AJ}             
\def\araa{ARA\&A}       
\def\apj{ApJ}           
\def\apjl{ApJ}          
\def\apjs{ApJS}         
\def\aap{A\&A}          
\def\aaps{A\&AS}        
\def\mnras{MNRAS}       

\begin{document}
\label{firstpage}
\maketitle

\begin{abstract}
We analysed two-dimensional maps of 48 early-type galaxies obtained
with the SAURON and OASIS integral-field spectrographs using
kinemetry, a generalisation of surface photometry to the higher order
moments of the line-of-sight velocity distribution (LOSVD). The maps
analysed include: reconstructed image, mean velocity, velocity
dispersion, $h_3$ and $h_4$ Gauss-Hermite moments. Kinemetry is a good
method to recognise structures otherwise missed by using surface
photometry, such as embedded disks and kinematic sub-components. In
the SAURON sample, we find that 31\% of early-type galaxies are single
component systems. 91\% of the multi-components systems have two
kinematic subcomponents, the rest having three. In addition, 29\% of
galaxies have kinematically decoupled components, nuclear components
with significant kinematic twists. We differentiate between slow and
fast rotators using velocity maps only and find that fast rotating
galaxies contain disks with a large range in mass fractions to the
main body. Specifically, we find that the velocity maps of fast
rotators closely resemble those of inclined disks, except in the
transition regions between kinematic subcomponents. This deviation is
measured with the kinemetric $k_5/k_1$ ratio, which is large and noisy
in slow rotators and about 2\% in fast rotators. In terms of E/S0
classification, this means that 74\% of Es and 92\% of S0s have
components with disk-like kinematics. We suggest that differences in
$k_5/k_1$ values for the fast and slow rotators arise from their
different intrinsic structure which is reflected on the velocity
maps. For the majority of fast rotators, the kinematic axial ratios
are equal to or less than their photometric axial ratios, contrary to
what is predicted with isotropic Jeans models viewed at different
inclinations. The position angles of fast rotators are constant, while
they vary abruptly in slow rotators.  Velocity dispersion maps of
face-on galaxies have shapes similar to the distribution of
light. Velocity dispersion maps of the edge-on fast rotators and all
slow rotators show differences which can only be partially explained
with isotropic models and, in the case of fast rotators, often require
additional cold components. We constructed local (bin-by-bin) $h_3 -
V/\sigma$ and $h_4 - V/\sigma$ diagrams from SAURON observations. We
confirm the classical anti-correlation of $h_3$ and $V/\sigma$, but we
also find that $h_3$ is almost zero in some objects or even weakly
correlated with $V/\sigma$. The distribution of $h_4$ for fast and
slow rotators is mildly positive on average. In general, fast rotators
contain flattened components characterised by a disk-like
rotation. The difference between slow and fast rotators is traceable
throughout all moments of the LOSVD, with evidence for different
intrinsic shapes and orbital contents and, hence, likely different
evolutionary paths.
\end{abstract}

\begin{keywords} galaxies: elliptical and lenticular - galaxies:
kinematics and dynamics - galaxies: structure - galaxies:evolution
\end{keywords}

%
%

\section{Introduction}
\label{s:intro}

The classification of galaxies both acknowledges the complexity of
these celestial objects and attempts to understand their formation and
evolution. The Hubble Classification of galaxies
\citep{1936RNeb..........H, 1961hag..book.....S,
  1975gaun.book.....S,1994cag..book.....S} recognises the dichotomy
between, broadly speaking, disk and ellliptical galaxies, for
historical reasons now often called late- and early-types. The
classification works well on the late-type galaxies in particular,
dividing the class into a number of subgroups which correlate with
properties such as bulge-to-disk ratio, morphology of spiral arms, gas
and dust content, to name a few, but it fails to bring a physical
insight to our understanding of early-types
\citep{1987IAUS..127..367T}, where the classification is based on
apparent shape and thus dependant on viewing angles.

In an effort to eliminate the unsatisfactory situation
\citet{1996ApJ...464L.119K} proposed a revision of the Hubble
classification. It was based on two discoveries in the 1970s and
1980s, both enabled by an improvement in the technical capabilities of
astronomical instruments. A series of papers
\citep{1975ApJ...200..439B, 1977ApJ...218L..43I} showed that bright
elliptical galaxies do not rotate as fast as they should, if they were
oblate isotropic systems supported by rotation
\citep{1978MNRAS.183..501B}, whereas less bright and generally smaller
systems, including also bulges of spirals, generally agree with such
predictions \citep{1983ApJ...266...41D, 1982modg.proc..113K,
  1982ApJ...256..460K}. A complementary discovery
\citep{1988A&A...193L...7B, 1988A&AS...74..385B,1989A&A...217...35B}
that the fast rotating galaxies are more likely to have disky
isophotes, while the slow rotating galaxies have boxy isophotes,
linked again the kinematics and shape of
galaxies. \citet{1996ApJ...464L.119K} changed the uniformity of
early-type galaxies to a dichotomy (disky vs boxy, fast vs slow,
brighter vs less bright) and linked the whole Hubble sequence from
right to left, from Sc--Sb--Sa to S0--E types, where 'rotation
decreases in dynamical importance compared to random motions'.

This important step forward introduced a readily measurable parameter
related to some physical properties. However, the higher order
variations in the isophotal shape (diskiness/boxiness) are not
measurable at all inclinations regardless of the prominence of the
disks \citep{1990ApJ...362...52R}, and finally, they are used to infer
the dynamical state of the galaxy. This might be a decent
approximation, especially if one assumes that all fast rotating
galaxies comprise of spheroidal slow rotating components and disks
seen at different inclination \citep{1999ApJ...513L..25R}, but the
edge-on observations of spheroidal components in spiral galaxies
showed that bulges are rotating fast as well
\citep{1982ApJ...256..460K}. To complicate things further the updated
classification continues to distinguish S0s from Es, keeping a viewing
angle dependent definition of S0s \citep{1990ApJ...348...57V}.

The choice of using the fourth order Fourier term in the
isophotometric analysis for classification is natural, because {\it
  (i)} it is much easier to take images of galaxies than to measure
their kinematics, and {\it (ii)} until recently it was not realistic
to spectroscopically map their two-dimensional structure.  The advent
of panoramic integral-field units, such as SAURON
\citep{2001MNRAS.326...23B}, is changing the technical possibilities
and the field itself; it is now possible to systematically map
kinematics of nearby galaxies up to their effective radii. We have
observed 72 nearby E, S0 and Sa galaxies as part of the SAURON survey
\citep[][hereafter Paper II]{2002MNRAS.329..513D}. Focusing here on a
subsample of 48 early-type galaxies (E, S0), these observations
clearly show the previously hinted rich variety of kinematic
substructures such as: kinematically decoupled cores, kinematic
twists, counter-rotating structures and central disks
\citep[][hereafter Paper III]{2004MNRAS.352..721E}.

Analysing the global properties of the SAURON velocity and velocity
dispersion maps, \citet[][hereafter Paper IX]{2007MNRAS.379..401E}
were able to separate the early-type galaxies into two physically
distinct classes of {\it slow} and {\it fast} rotators, according to
their specific (projected) angular momentum measured within one
effective radius, $\lambda_R$. This finding augments the view that led
to the revision of the classification, but the SAURON observations
provide the crucial quantitive data. Moreover, the results of Paper IX
suggest a way to dramatically improve on the Hubble classification and
substitute S0s and (misclassified) disky ellipticals with one class of
fast rotators.

\citet[][hereafter Paper X]{2007MNRAS.379..418C} addressed again the
issue of orbital anisotropy of early-type galaxies. They constructed
the ($V/\sigma$, $\epsilon$) diagram \citep{1978MNRAS.183..501B} using
an updated formalism \citep{2005MNRAS.363..937B}, and compared it with
the results from general axisymmetric dynamical models for a subsample
of these galaxies \citep[][hereafter Paper
  IV]{2006MNRAS.366.1126C}. They found that slow and fast rotators are
clearly separated on the ($V/\sigma$, $\epsilon$) diagram (unlike Es
and S0s), such that slow rotators are round, moderately isotropic and
are likely to be somewhat triaxial, while fast rotators appear
flattened, span a larger range of anisotropies, but are mostly oblate
axisymmetric objects. This finding is in a partial agreement with
previous studies which either found round early-type galaxies radially
anisotropic \citep{1991MNRAS.253..710V}, moderately radially
anisotropic \citep{2001AJ....121.1936G} or only weakly anisotropic
with a range of anisotropies for flattened systems
\citep{2003ApJ...583...92G}. The results of Paper X, however, clearly
show that intrinsically flatter galaxies tend to be more anisotropic
in the meridional plane. The models also indicate that the fast
rotators are often two component systems, having also a flat and
rotating, kinematically distinct, disk-like component.

Dynamical models are often time consuming and difficult to
construct. Ultimately, one would like to be able to classify galaxies
by their observable properties only. Is it possible to learn about the
intrinsic shapes of the early-type galaxies from observations only?
Surface photometry, being but the zeroth moment of the ultimate
observable quantity for distant galaxies, the line-of-sight velocity
distribution (LOSVD), cannot give the final answer. It is necessary to
look at the other moments of the LOSVD: mean velocity, velocity
dispersion and higher-order moments, commonly parameterised by
Gauss--Hermite coefficients, $h_3$ and $h_4$
\citep{1993ApJ...407..525V, 1993MNRAS.265..213G}, which measure
  asymmetric and symmetric deviation of the LOSVD from a Gaussian,
  respectively.

  Indeed, in the last dozen years several studies investigated higher
  moments of the LOSVD of early-type galaxies observing them along one
  or multiple slits
  \citep[e.g.][]{1994MNRAS.269..785B,1994MNRAS.268..521V,
    1994MNRAS.270..271V,2000A&AS..144...53K,
    2000A&AS..145...71K,2001MNRAS.326..473H,2002A&A...395..753W,
    2006MNRAS.371..633H,2008ApJS..175..462C}. These studies deepened
  the dichotomy among early-type galaxies showing that fast rotating
  galaxies with disky isophotes also exhibit an anti-correlation
  between $h_3$ and $V/\sigma$. This is consistent with these galaxies
  being made of two components: a bulge and a disk. The symmetric
  deviations, on the other hand are usually smaller than asymmetric
  ones, and somewhat positive in general. In addition, the observed
  higher order moments of the LOSVD can be used to constrain the
  possible merger scenarios of early-type galaxies and their formation
  in general \citep[e.g.][]{1991A&A...249L...9B,
    2000MNRAS.316..315B,2006MNRAS.372L..78G,2006MNRAS.372..839N,2007MNRAS.376..997J}. However,
  observations along one or two slits are often not able to describe
  the kinematical richness of early-type galaxies.

In this work we use \kinemetry~\citep{2006MNRAS.366..787K}, a
generalisation of surface photometry to all moments of the LOSVD, to
study SAURON maps of 48 early-type galaxies. The purpose of this work
is to investigate observational clues from resolved two-dimensional
kinematics for the origin of the differences between the slow and fast
rotators.

In Section~\ref{s:dat} we briefly remind the reader of the SAURON
observations and data reduction. Section~\ref{s:met} describes the
methods and definitions used in this work. The main results are
presented in Section~\ref{s:res}. In Section~\ref{s:disc} we offer an
interpretation of the results and we summarise the conclusions in
Section~\ref{s:conc}. In Appendix~\ref{s:psf} we discuss the influence
of seeing on the two-dimensional kinematics and in
Appendix~\ref{s:profiles} we present the radial profiles of the
kinemetric coefficients used in this study.

%
%

\section{Sample and Data}
\label{s:dat}

In this work we used the data from the SAURON sample which was
designed to be {\it representative} of the galaxy populations in the
plane of ellipticity, $\epsilon$, versus absolute $B$ band magnitude
$M_B$. The sample and its selection details are presented in Paper II.
In this study we focus on the 48 galaxies of the SAURON E+S0 sample.

SAURON is an integral-field spectrograph with a field-of-view (FoV) of
about $33\arcsec \times 41\arcsec$ and $0\farcs94 \times 0\farcs94$
square lenses, mounted at the William Herschel
Telescope. Complementing the SAURON large scale FoV, we probed the
nuclear regions of a number of galaxies with OASIS, then mounted at
Canada-Hawaii-France Telescope, a high spatial resolution
integral-field spectrograph, similarly to SAURON based on the TIGER
concept \citep{1995A&AS..113..347B}. The FoV of OASIS is only
$10\arcsec \times 8\arcsec$, but the spatial scale is $0\farcs27
\times 0\farcs27$, fully sampling the seeing-limited point spread
function and providing on average a factor of 2 improvement in spatial
resolution over SAURON. The spectral resolution of OASIS is, however,
about 20\% lower than that of SAURON, and only a sub-sample of the
SAURON galaxies was observed.

In this paper we are investigating the stellar kinematics of
early-type galaxies. Paper III and \citet[][hereafter Paper
  VIII]{2006MNRAS.373..906M} discuss the extraction of kinematics and
construction of maps of the mean velocity $V$, the velocity dispersion
$\sigma$, and the Gauss-Hermite moments $h_3$ and $h_4$ in great
detail. All maps used in this work are Voronoi binned
\citep{2003MNRAS.342..345C} to the same signal-to-noise ratio. The
SAURON kinematic data used here are of the same kinematic extraction
as in Paper X with the latest improvement on the template mismatch
effects in higher moments of the LOSVD. The SAURON mean velocity maps
are repeated in this paper for the sake of clarity, but we encourage
the reader to have copies of both Paper III and Paper VIII available
for reference on other moments of the LOSVD.

%
%

\section{Method and definitions}
\label{s:met}

Maps of the moments of the LOSVD offer a wealth of information, but
also suffer from complexity. It is difficult, if not impossible, to
show error bars for each bin on the map, and the richness of the maps
can lead to the useful information being lost in detail. As in the
case of imaging, it is necessary to extract the useful information
from the maps to profit from their two-dimensional coverage of the
objects. In this section we describe the method used to analyse the
maps and discuss definitions utilised throughout the paper.

\subsection{Kinemetry}
\label{ss:kin}

\citet{2006MNRAS.366..787K} presented kinemetry, as a quantitative
approach to analysis of maps of kinematic moments.  Kinemetry is a
generalisation of surface photometry to the higher-order moments of
the LOSVD. The moments of the LOSVD have odd or even parity. The
surface brightness (zeroth moment) is even, the mean velocity (first
moment) is odd, the velocity dispersion (second moment) is even,
etc. Kinemetry is based on the assumption that for the odd moments the
profile along the ellipse satisfies a simple cosine law, while for the
even moments the profile is constant (the same assumption is also used
in surface photometry). Kinemetry derives such best-fitting ellipses
and analyses the profiles of the moments extracted along these by
means of harmonic decomposition. It follows from this that the
application of kinemetry on even moments is equivalent to surface
photometry resulting in the same coefficients for parameterisation of
the structures (e.g position angle, ellipticity and fourth order
harmonics).

Application of kinemetry on odd maps such as velocity maps\footnote{It
  is customary in the literature to refer to the maps of mean velocity
  as {\it velocity fields}. Sometimes, due to the specific shape of
  contours of constant velocities, velocity maps are referred to as
  {\it spider diagrams}
  \citep[e.g.][]{1978ARA&A..16..103V}. Two-dimensional representations
  of the next moment are, however, usually referred to as velocity
  dispersion maps. Instead of alternating between {\it fields} and
  {\it maps} we choose to call all two-dimensional representations of
  the LOSVD maps: velocity map, velocity dispersion map, $h_3$
  map....} provides radial profiles of the kinematic position angle
\PAk, axis ratio or flattening, \qk=b/a (where b and a are lengths of
minor and major axis, respectively), and odd harmonic terms obtained
from the Fourier expansion (since velocity is an odd map, even terms
are, in principle, not present, while in practice are very small and
usually negligible). In the case of stellar velocity maps, the
dominant kinemetry term is $k_1=\sqrt{a_1^2+b_1^2}$, representing the
velocity amplitude, where $a_1$ and $b_1$ are the first sine and
cosine terms, respectively.  The deviations from the assumed simple
cosine law are given by the first higher order term that is not
fitted, $k_5=\sqrt{a_5^2+b_5^2}$, usually normalised with respect to
$k_1$. These four parameters form the basis of our analysis because
they quantify the kinematical properties of the observed galaxies:
orientation of the map (a projection of the angular momentum), opening
angle of the iso-velocity contours, the amplitude of the rotation and
the deviation from the assumed azimuthal variation of the velocity
map. For the other moments of the LOSVD one could derive similar
quantities, depending on the parity of the moment. As will be
discussed below, we focus on kinemetry coefficients that describe
velocity maps in detail and some specific kinemetry coefficients from
the maps of the higher order moments. A detailed description of the
method, error analysis and parameters is given in
\citet{2006MNRAS.366..787K}.

\subsection{Radial profiles}
\label{ss:prof}

Kinemetric radial profiles can be obtained along ellipses of different
axial ratios and position angles. At each radius there is the best
fitting ellipse, along which a profile of the kinematic moment will
have a certain shape: it follows a cosine or it is constant, for odd
and even moments, respectively. If this is the case, the higher order
Fourier terms are non-existent or at least negligible for such an
ellipse.

In the case of even moments, the best fitting ellipses describe the
underlying iso-contours, like isophotes in the case of surface
photometry, or contours of constant velocity dispersion, iso-$\sigma$
contours. In the case of odd moments, this is somewhat more difficult
to visualise, but the axial ratio of the best-fitting ellipse is
related to the opening angle of the iso-velocity contours: the larger
the axial ratio, the more open is the spider diagram of the velocity
map.

In this study, kinemetry is used for extraction of parameters in the
following ways:

\begin{itemize} 

\item[{\it i)}] We apply kinemetry to SAURON reconstructed images of
  galaxies, which are obtained by summing the spectra along the
  spectral direction at each sky position. This is equivalent to low
  resolution surface photometry on galaxies from the sample. We focus
  on the photometric position angle \PAp~and photometric axial ratio,
  related to ellipticity as \qp=1-$\epsilon$. In this case, kinemetry
  is used in its {\it even} mode, where even harmonics are fitted to
  the profiles extracted along the best fitting ellipses.

\item[{\it ii)}] We use kinemetry to derive radial profiles of the four
  parameters that describe velocity maps: \PAk, \qk, $k_1$ and
  $k_5$. In this case, kinemetry is applied to the maps in its {\it
    odd} mode, when only odd Fourier harmonics are fit to the profiles
  extracted along the best fitting ellipses, which are, in general,
  different from the best fitting ellipse of {\it (i)}. In some cases
  when it is not possible to determine the best-fitting ellipse we run
  kinemetry on circles (see below).

\item[{\it iii)}] Kinemetry is applied to velocity dispersion maps,
  using the {\it even} mode as in {\it (i)}. In this case, however,
  the parameters of the ellipses used to extract profiles were fixed
  to the best fitting values of surface photometry obtained in {\it
    (i)}.

\item[{\it iv)}] Maps of Gauss-Hermite coefficients $h_3$ and $h_4$
  were also parameterised using kinemetry in {\it odd} and {\it even}
  mode, respectively. In both cases, we used the best fitting ellipses
  from the lowest {\it odd} (velocity map) and {\it even} moment
  (reconstructed image), respectively.

\end{itemize}

Before proceeding it is worth explaining in more detail our decision
not to use kinemetry to fit the ellipses in some cases. Under {\it
  (ii)} we mentioned that on some velocity maps it was necessary to
run kinemetry on circles. In general, the mean stellar velocity has an
odd parity, and its map, in an inclusive triaxial case, will be
point-antisymmetric. Certain maps, however, do not follow this rule,
having no detectable net rotation, e.g. NGC4486, or the inner part of
NGC4550. In the latter case, the zero velocity in the inner part can
be explained by the superposition of two counter-rotating stellar
components as advocated by \citet{1992ApJ...394L...9R} and
\citet{1992ApJ...400L...5R}, where the mass of the counter-rotating
component is about 50\% of the total mass (Paper X). In other cases
the non rotation could be a result of dominant box orbits which have
zero angular momentum. The basic assumption of kinemetry for odd
kinematic moments therefore breaks down resulting in velocity maps
that appear noisy and one cannot expect reasonable results.

In practice, this means that the best-fitting ellipse parameters for
maps without net rotation will not be robustly determined (degeneracy
in both \PAk and \qk) while the higher harmonic terms will be large
and meaningless. Specifically, $k_5$ will have high values. We
partially alleviate this degeneracy by first running an unconstrained
kinemetry fit on stellar velocity maps and identifying maps where
$k_5/k_1 > 0.1$ and corresponding radii where it occurs. At these
radii we re-run kinemetry, but using circles for extraction of
velocity profiles and Fourier analysis. In this way we set the axial
ratio \qk=1 in order to break the degeneracy. Although the $k_5/k_1$
term cannot be directly compared with the $k_5/k_1$ term obtained from
a best-fitting ellipse, in this case, if there is any indication of
odd parity in the map, we can still determine the local amplitude of
rotation $k_1$, and give a good estimate for \PAk.

The other note refers to items {\it (iii)} and {\it (iv)}. Although,
in principle, it would be possible to run kinemetry freely on the
velocity dispersion maps, or maps of higher Gauss-Hermite moments
(e.g. $h_3$ and $h_4$), the noise in the data is too high to give
trustworthy results for the whole sample. By setting the shape of the
curve to the best-fitting ellipses of the corresponding lowest odd or
even moment, the harmonic terms of kinemetry quantify the differences
between these even and odd moments of the LOSVD.

\begin{figure}
        \includegraphics[width=\columnwidth, bb=30 30 590 590]{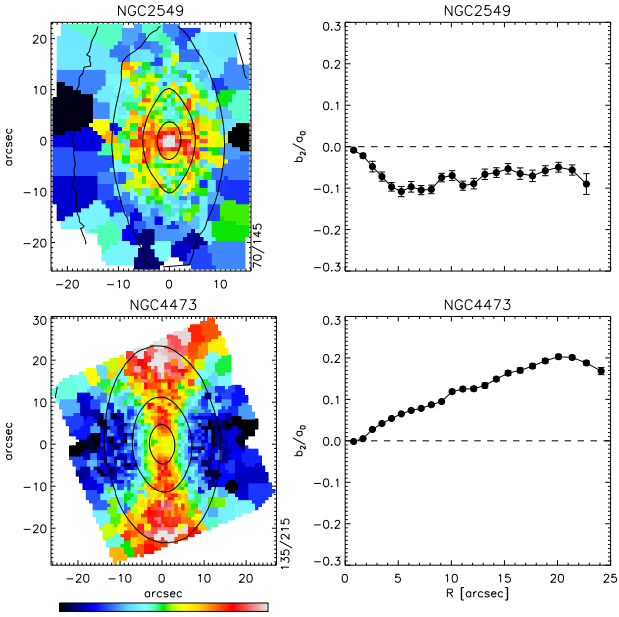}
  \caption{\label{f:diff} {\bf Left:} Voronoi binned stellar velocity
    dispersion maps of NGC2549 (top) and NGC4473
    (bottom). Over-plotted lines are isophotes. In the lower right
    corner of each map are values that correspond to minimum and
    maximum colours on the colour bar. {\bf Right:} Radial profiles of
    $b_2/a_2$ kinemetry terms for NGC2549 (top) and NGC4473 (bottom).}
\end{figure}

An example of expected differences can be visualised comparing the
isophotes of the surface brightness and the stellar velocity
dispersion maps of NGC2549 and NGC4473 (Fig.~\ref{f:diff}). On both
images isophotes are aligned with the vertical (y) axis of the
maps. In the case of NGC2549 the contours of constant velocity
dispersion seem to be perpendicular to the isophotes, at least within
the central 10\arcsec, while in the rather unusual case of NGC4473, the
high values of velocity dispersion have the same orientation as the
isophotes. The physical explanation of these striking differences
should be looked for in the internal orbital structure. We postpone
this discussion to Section~\ref{ss:shape}.

The noise and irregular shape of iso-$\sigma$ contours decrease the
usefulness of fitting for the velocity dispersion contours. Extracting
harmonic terms along the isophotes, however, can yield a clear signal
of the different shape of these two moments. An extracted velocity
dispersion profile in these two cases (for example along the second
brightest isophote shown on Fig.~\ref{f:diff}) will go through two
maxima and two minima. The minima and maxima of these two profiles
will be out of phase, because along the major axis in NGC2549 there is
a decrease in velocity dispersion while in NGC4473 there is an
increase. The decomposition of these two profiles will give different
amplitudes to the harmonic terms. Specifically, $b_2$ (cosine) term
will be the most influenced, because this term is related to the error
in the axial ratio \citep{1987MNRAS.226..747J}, and the shapes such as
in NGC2549 and NGC4473 will give negative (iso-$\sigma$ rounder than
isophotes) and positive (iso-$\sigma$ flatter than isophotes) $b_2$,
respectively. An alternative way to visualise the difference between
these two maps based on the values of $b_2$ is to consider that a
negative $b_2,$ corresponds to a decrease of the $\sigma$ at the major
axis of the best fitting ellipse compared to the $\sigma$ measured at
the minor axis of the best fitting ellipse (NGC2549).  In contrast, a
positive $b_2$ corresponds to an increased value of the $\sigma$ at
the major axis position of the best fitting ellipse (NGC4473). By
monitoring these harmonic terms it is possible to quantify the shape
difference between the observed zeroth and second moments of the
LOSVD.

\subsection{Definition of structures on stellar velocity maps}
\label{ss:def_comp}

A few kinemetric profiles are able to describe a wealth of information
from the maps. Specifically, we wish to use them to highlight the
kinematic structures on the maps and to recognise hidden kinematic
components. Here we present a set of quantitative criteria for
describing features on the stellar velocity maps. Some of the criteria
are dependent on the quality of the data and they should be modified
if used on maps obtained with other IFUs. The following rules were
presented by \citet{2006MNRAS.366..787K} and Paper IX, but here we
list them for the sake of clarity.

A single velocity map can contain a number of kinematic
components. Often they are easily recognisable by visual
inspection. In a quantitive way we differentiate between:

\begin{itemize}

\item {\it Single Component} (SC) map: having a radially constant or slow
varying \PAk and \qk profiles.

\item {\it Multiple Component} (MC) map: characterised with an abrupt
change in either: $\Delta$\qk$ > 0.1$, or $\Delta$\PAk $> 10\degr$,
or a double-hump in $k_1$ with a local minimum in between, or a peak
in $k_5$ where $k_5/k_1>0.02$.

\end{itemize}

MC maps are clearly more complex than SC maps. The above values for
changes to the kinemetric coefficients are used to determine the
extent of each subcomponent (components $C_1$, $C_2$ and $C_3$ with
radii $R_{12}$ and $R_{23}$ between them). Each subcomponent can be
described as being of the following type (limiting values apply for
the SAURON dataset):

\begin{itemize}

\item {\it Disc-like rotation} (DR): defined when the higher order
  harmonic $k_5/k_1 < 0.02$, while the variation of \qk~and \PAk~ is
  less than 0.1 and $10\degr$, respectively. Note that this name {\it
    does not} imply that the object is a disk intrinsically .

\item {\it Low-level velocity} (LV): defined when the maximum of $k_1$
  is lower than $15 \kms$. A special case is {\it Central low-level
    velocity} (CLV) when LV occurs in the central kinematical
  component on the map.

\item {\it Kinematic misalignment} (KM): defined when the absolute
  difference between the photometric \PAp~and kinemetric position
  \PAk~angles is larger than $10\degr$.

\item {\it Kinematic twist} (KT): defined by a smooth variation of the
  kinematics position angle \PAk~with an amplitude of at least
  $10\degr$ within the extent of the kinematic component.

\item {\it Kinematically decoupled component} (KDC): if there is an
  abrupt change in \PAk, with a difference larger than $20\degr$
  between two adjacent components, or if there is an outer LV
  component (in which case the measurement of \PAk~is uncertain). A
  special case of KDCs are {\it Counter rotating cores} (CRC) where
  $\Delta $\PAk~between the components is $180\degr$ (within the
  uncertainties).

\end{itemize}

Most of the above definitions are new, arising from two-dimensional
maps which offer a more robust detection of structures. The definition
of KDC is, however, similar to the one used in the past
\citep[e.g.][]{1988A&A...202L...5B,1991AJ....102..882S}, where the
two-dimensional coverage enables a determination of the orientation of
the kinematic components. It should be noted that classification of a
kinematic component as a CLV is strongly dependent on the spatial
resolution of the instrument. As will be seen later, higher spatial
resolution can change the appearance and therefore the classification
of the components.

Similarly, it should be stressed that the limiting values used for
these definitions are geared towards the SAURON data. The OASIS data,
due to different instrumental properties and observing set-up, will
have somewhat different limiting values, mostly arising in the higher
order Fourier terms. For example, the mean uncertainty of $k_5/k_1$
term for the OASIS sample is 0.033, significantly higher compared to
the one for the SAURON sample (0.015). In order to treat consistently
the two data sets, we adopt a somewhat more conservative value of 0.04
as the limiting values for $k_5/k_1$ in definition of DR component
when estimated from the OASIS data.

While abrupt changes in the orientation, axial ratio, or velocity
amplitude are intuitively clear as evidences for separate kinematic
components, the $k_5/k_1$ as an indicator of components is more
complex to comprehend. Still, the simple models of two kinematic
components rotating at a given relative orientation give rise to
$k_5/k_1$ term in the kinemetric expansion in the region where these
components overlap \citep{2006MNRAS.366..787K}. Since we measure
luminosity weighted velocities, the position and extent of the raised
$k_5/k_1$ region depends on relative luminosity contributions of the
components, marking the transition radii between the components and
not their start or end. Furthermore, it is also, necessary, to
distinguish between high $k_5/k_1$ due to a super-position of
kinematic components (a genuine signal) and high $k_5/k_1$ originating
from noisy maps, such as maps with no detectable rotation
(e.g. NGC4486) or large bin-to-bin variations (e.g. OASIS map of
NGC3379). For our data, when the signal in $k_5$ is $10\%$ of $k_1$,
we consider the noise too high and the $k_5/k_1$ ratio not usable for
detecting individual components.

\subsection{Seeing and quantification of kinematic components}
\label{ss:skc}

Robust estimates of the number of sub-components in velocity maps and
their sizes are influenced by three major factors: data quality,
physical properties and seeing. While the data quality is described by
measurement uncertainties, and in that sense it is quantifiable to
some extent, the other two factors are more complex. By 'physical
properties' we assume physical processes that hide kinematic
information from our view, such as specific orientation of the object,
dust obscuration or simply the fact that we are measuring luminosity
weighted quantities and we might miss kinematic components made up of
stars that constitute a low luminosity fraction of the total
population.

The influence of seeing is particularly relevant for subcomponents in
the centres of galaxies. In Appendix~\ref{s:psf} we tested the
dependence of the kinemetric coefficients on representative seeings,
for velocity maps viewed at different orientations. This exercise
showed that: {\it(i)} \PAk~and $k_5/k_1$ are not significantly
influenced by the seeing, {\it (ii)} the amplitude and, to a minor
extent, the shape of $k_1$ are somewhat influenced by the seeing, and
{\it (iii)} the axial ratio \qk~can be strongly influenced by the
seeing (Fig.~\ref{f:prof_fwhm}). In addition to these conclusions, the
test showed that the inclination of an object is also a factor
contributing to the change of the intrinsic \qk, and to a minor
extent, $k_1$ profiles, where higher inclinations are particularly
influenced by the seeing effects.

In practice, this means that the change in \qk~is a less robust
indicator of kinematic components. We found that more robust
indicators are abrupt changes in $k_5/k_1$ and \PAk~profiles, double
humps in $k_1$ profiles or decrease of $k_1$ amplitude below our
detection limit for rotation. We used these as estimates of the sizes
of kinematic components. It should be, however, noted that the size of
a component is just a luminosity weighted estimate, originating from a
super-position of luminosities of individual components, and the
component can intrinsically extend beyond that radius. Only detailed
dynamical models could give a more robust estimate of the internal
orbital structure.

\subsection{Determination of global and average values}
\label{ss:glob}

In addition to radial profiles we present in this paper a number of
average quantities. Similar luminosity-weighted quantities have
already been derived in Papers IX and X: global \PAk, global \PAp,
average $\epsilon$. In this study we use the velocity maps to
determine the luminosity weighted average $\langle$\PAk$\rangle$,
$\langle$\qk$\rangle$, and $\langle k_5/k_1 \rangle$ for the whole map
and for each kinematic component. We also measured the luminosity
weighted $\langle$\PAp$\rangle$, $\langle$\qp$\rangle$ from the
reconstructed images (both global and for each component), $\langle
b_2/a_0 \rangle$ from velocity dispersion maps and average values of
$h_4$ (measured as the $a_0$ harmonic term) from $h_4$ maps. In
practice, we do this following the expression from Paper IX. The mean
$\langle G \rangle$ of a quantity $G(R)$ derived from its sampled
radial profiles can be approximated with:
\begin{equation}
\label{eq:av}
\langle G \rangle \sim \frac{ \sum_{k=1}^{N} q(R_k) F(R_k) G(R_k) (R^2_{out,k} - R^2_{in,k}) }{\sum_{k=1}^{N}q(R_k) F(R_k) (R^2_{out,k} - R^2_{in,k})}
\end{equation}
\noindent where $q(R_k)$ and $F(R_k)$ are the axial ratio and the
surface brightness of the best fit ellipse, with semi-major axis
$R_k$.  Eq.~(\ref{eq:av}) is based on an expression defined in
\citet{1999ApJ...517..650R}. The uncertainties of these average values
are calculated in the standard way as the sum of the quadratic
differences between the average value and the value at each position
$R_k$.

In Paper X the global \PAk~was derived using the formalism from
Appendix C in \citet{2006MNRAS.366..787K}. This approach differs from
the one described here in the sense that it is less sensitive to the
kinematic structures in the central region, such as abrupt changes of
PA in case of a KDC. That approach is well suited for making global
comparisons between \PAk~or \PAp, such as global kinematic
misalignment, when it is required that they are measured on large
scales to avoid influence of local perturbations in the nuclear
regions (e.g. seeing, dust, bars). In this study, however, we want to
compare the radial properties of different moments of the LOSVD and
for that reason we use the approach of Paper IX to all measured
quantities. Note that for the purpose of the direct comparison we
measured both kinematic and photometric quantities on the SAURON data,
in contrast with Papers IX and X.

%
%

\section{Results}
\label{s:res}

In this section we present the results of kinemetric analysis of the
LOSVD moments maps. We look at the presence of kinematic substructures
in velocity maps (Section~\ref{ss:subcomp}), properties of radial
profiles of \PAk, $k_1$, $\sigma$ (Section~\ref{ss:SRFR}), comparison
between \qk~and \qp~(Section~\ref{ss:flat}), the shape difference
between isophotes and iso-$\sigma$ contours (Section~\ref{ss:shape})
and properties of $h_3$ and $h_4$ Gauss-Hermite moments
(Section~\ref{ss:high}) with the purpose to investigate the internal
structure of SAURON galaxies. Kinemetry probes local characteristics
of galaxies, and we wish to link those with the global properties
described in Papers IX and X. In this analysis, the most useful is the
first moment of the LOSVD, the mean velocity, because it is a moment
rich in structure and with the strongest signal. We present the
kinemetric profiles of this moment in
Appendix~\ref{s:profiles}. Although kinemetry is performed on other
moments of the LOSVD, we discuss the dominant terms only.

\subsection{Substructures on the velocity maps}
\label{ss:subcomp}

\begin{table*}
\caption{Kinemetric properties of the 48 E and S0 {\tt SAURON}
  galaxies.}
\label{t:kin_prop}
\begin{tabular}{llcrrccclcl}
\hline
Galaxy & Group & $N_C$ & $R_{12}$ & $R_{23}$ & $C_1$ & $C_2$ & $C_3$ & $KM$ & Rotator &Comment \\
 ~~~(1) & ~~(2) & (3) & (4) & (5) & (6) & (7) & (8) & (9) & (10) &(11) \\
\hline
\noalign{\smallskip}
NGC0474 &  MC       & 2 & 7    & --   & DR & DR &  -- & KM(1,2)  & F & KT between $C_1$ and $C_2$\\ 
NGC0524 &  SC       & 1 & --   & --   & DR & -- &  -- & --       & F & Possible $C_2$ beyond r=12\arcsec\\ 
NGC0821 &  SC       & 1 & --   & --   & DR & -- &  -- & --       & F & Flat $k_1$ profile \\
NGC1023 &  SC       & 1 & --   & --   & KT & -- &  -- & --       & F & $k_5/k_1< 0.02$,\\ 
NGC2549 &  MC       & 2 & 13   & --   & DR & DR &  -- & --       & F & $C_2$: Flat $k_1$ profile \\
NGC2685 &  SC       & 1 & --   & --   & DR & -- &  -- & --       & F & --\\ 
NGC2695 &  MC       & 2 & 7    & --   & DR & DR &  -- & --       & F & -- \\  
NGC2699 &  MC       & 2 & 6    & --   & DR & DR &  -- & KM(1,-)  & F & $C_2$: Flat $k_1$ profile   \\ 
NGC2768 &  SC       & 1 & --   & --   & -- & -- &  -- & --       & F &$r \lesssim 10\arcsec$ rigid body rotation; $k_5/k_1<0.02$ for $r>10\arcsec$\\
NGC2974 &  SC       & 1 & --   & --   & DR & -- &  -- & --       & F & -- \\ 
NGC3032 &  MC (CLV) & 2 & 2.5  & --   & LV & DR &  -- & --       & F & CRC in OASIS map\\  
NGC3156 &  SC       & 1 & --   & --   & DR & -- &  -- & --       & F & --\\
NGC3377 &  SC       & 1 & --   & --   & KT & -- &  -- & --       & F & $k_5/k_1 < 0.02$ over the map  \\  
NGC3379 &  SC       & 1 & --   & --   & DR & -- &  -- & --       & F & --\\
NGC3384 &  MC       & 2 & 10   & --   & DR & DR &  -- & KM(-,2)  & F & --  \\ 
NGC3414 &  MC (KDC) & 2 & 10   & --   & DR & LV &  -- & KM(-,2*) & S & CRC  \\ 
NGC3489 &  MC       & 2 & 6    & --   & DR & DR &  -- & --       & F & -  \\ 
NGC3608 &  MC (KDC) & 2 & 10   & --   & DR & LV &  -- & KM(-,2*) & S & CRC  \\
NGC4150 &  MC (CLV) & 3 & 3.5  & 9.5  & LV & -- &  DR & KM(1,-,-)& F & KDC in OASIS map with $r_{size}=1\farcs5$  \\ 
NGC4262 &  MC       & 2 & 9    & --   & DR & -- &  -- & KM(1,2)  & F & --  \\ 
NGC4270 &  MC       & 2 & 6    & --   & -- & -- &  -- & --       & F & Possible there components\\
NGC4278 &  MC       & 2 & 16   & --   & DR & -- &  -- & KM(-,2)  & F & $C_2$: Decreasing $k_1$ profile  \\ 
NGC4374 &  SC       & 1 & --   & --   & LV & -- &  -- & KM*      & S & -  \\ 
NGC4382 &  MC (CLV) & 3 & 2    & 14.5 & LV & DR &  DR & KM(1,-,-)& F & CRC in OASIS map, KT between $C_1$ and $C_2$  \\ 
NGC4387 &  MC       & 2 & 7    & --   & DR & DR &  -- & --       & F & Decreasing $k_1$ beyond $r=13\arcsec$  \\ 
NGC4458 &  MC (KDC) & 2 & 3    & --   & -- & LV &  -- & KM(-,2*) & S & --  \\ 
NGC4459 &  MC       & 2 & 12   & --   & DR & DR &  -- & KM(1,-)  & F & -- \\ 
NGC4473 &  MC       & 2 & 10   & --   & DR & -- &  -- & --       & F & $C_1$: possible KT. $C_2$: decreasing $k_1$ profile. \\ 
NGC4477 &  SC       & 1 & --   & --   & DR & -- &  -- & KM       & F & --\\
NGC4486 &  SC       & 1 & --   & --   & LV & -- &  -- & KM*      & S & --  \\ 
NGC4526 &  MC       & 2 & 11   & --   & DR & -- &  -- & --       & F & $C_2$: decreasing $k_1$ profile  \\ 
NGC4546 &  MC       & 2 & 9    & --   & DR & DR &  -- & KM(1,-)  & F & $C_2$: Flat $k_1$ profile \\ 
NGC4550 &  SC       & 1 & --   & --   & LV & -- &  -- & KM*      & S & Two co-spatial counter-rotating disks not detected  \\      
NGC4552 &  MC (KDC) & 2 & 4    & --   & KT & -- &  -- & KM(1,2)  & S & Flat $k_1=15\kms$ over the map\\ 
NGC4564 &  SC       & 1 & --   & --   & DR & -- &  -- & --       & F & -- \\ 
NGC4570 &  MC       & 2 & 8    & --   & DR & DR &  -- & --       & F & -- \\ 
NGC4621 &  MC (KDC) & 2 & 4    & --   & DR & DR &  -- & KM(1,-)  & F & CRC in OASIS with r$\sim1\farcs5$  \\ 
NGC4660 &  MC       & 2 & 7    & --   & DR & DR &  -- & --       & F & -- \\ 
NGC5198 &  MC (KDC) & 2 & 2.5  & --   & -- & KT &  -- & KM(1*,2*)& S & $C_2$: LV between 2.5-10\arcsec  \\
NGC5308 &  MC       & 2 & 7    & --   & DR & DR &  -- & --       & F & No signature of $C_2$ in $k_5/k_1$ \\ 
NGC5813 &  MC (KDC) & 2 & 12   & --   & DR & LV &  -- & KM(-,2*) & S & --  \\
NGC5831 &  MC (KDC) & 2 & 8    & --   & DR & LV &  -- & KM(-,2*) & S & --  \\
NGC5838 &  MC       & 2 & 6    & --   & DR & DR &  -- & --       & F & No signature of $C_2$ in $k_5/k_1$   \\
NGC5845 &  MC       & 2 & 4.5  & --   & DR & DR &  -- & --       & F & --  \\
NGC5846 &  SC       & 1 & --   & --   & LV & -- &  -- & KM*      & S & --  \\
NGC5982 &  MC (KDC) & 2 & 3.5  & --   & -- & LV &  -- & KM(1,-)  & S & $C_2$: LV is between 2-8\arcsec  \\
NGC7332 &  MC (KDC) & 3 & 3    & 12   & KT & DR &  DR & KM(1,-,-)& F & $C_2$: continuously increasing $k_1$; KDC only in PA change\\  
NGC7457 &  MC (CLV) & 2 & 3    & --   & KT & DR &  -- & KM(1,2)  & F & $C_2$: continuously increasing $k_1$ \\    
\hline
\end{tabular}
\\
Notes:
(1)~Galaxy identifier; 
(2)~Kinematic galaxy group: see text for details;
(3)~Number of kinematic components;
(4)~Transition radius between the first and second components (arcsec);
(5)~Transition radius between the second and third components (arcsec);
(6), (7) and (8)~Kinematic group for the first second and third components;
(9)~Local kinematic misalignment between luminosity-weighted averages of
   \PAk~and \PAp: numbers refer to the kinematic component and * notes
   that the \PAk~was determined in the region with $k_1$ mostly below
   $15 \kms$;
(10)~Rotator class: S -- slow rotator, F -- fast rotator
(11)~Comment.
\end{table*}

Looking at the kinemetric profiles of 48 SAURON galaxies
(Fig.~\ref{f:prof}) the following general conclusions can be made:

\begin{itemize}

\item \PAk~profiles are in general smooth and often constant
  (e.g. NGC2974) or mildly varying (e.g. NGC474). In some cases there
  are abrupt changes of up to $180\degr$ within $1-2\arcsec$
  (e.g. NGC3608).

\item Profiles of the axial ratio \qk~are generally smooth and often
  similar to \qp~profiles (e.g. NGC1023, NGC3384, NGC4570).

\item There is a variety of $k_1$ profiles, most of them rise and
  flatten, but some continue to rise, while some drop (e.g. NGC4278,
  NGC4477 and NGC4546).

\item Considering the $k_5/k_1$ radial dependence, there are three
  kinds of objects: those that have the ratio below 0.02 along most of
  the radius (e.g. NGC2974), those that have the ratio greater than
  0.1 along most of the radius (e.g. NGC4374) and those that have the
  ratio below 0.02 with one (or more) humps above this value
  (e.g. NGC2549).

\item Objects with $k_5/k_1 > 0.1$ along a significant part of the
  profile from SAURON data are all classified as slow rotators in
  Paper IX (e.g. NGC4486).

\end{itemize}

A further step in understanding the complex velocity maps can be made
by applying definitions of kinemetric groups (see
Section~\ref{ss:def_comp}) to the radial profiles. They are summarised
in Table~\ref{t:kin_prop}. There are 15 galaxies characterised as SC
(31\%)\footnote{Although NGC4550 is made of two counter-rotating and
  co-spatial disks (Section~\ref{ss:prof}), it is formally
  characterised as a SC galaxy due to its low velocity within the
  SAURON FoV.}, the rest being MC galaxies (69\%) of which 10 have KDC
(21\%) and 4 CLV (8\%). Higher resolution observations with OASIS,
however, show that all SAURON CLVs are in fact small KDCs and,
moreover, CRCs (Paper VIII). This means that there are actually 14
(29\%) KDCs in the SAURON sample. Kinematic profiles of the OASIS data
also clearly show structures that are partially resolved in the SAURON
observations, such as KDC (NGC4621, NGC5198, NGC5982) or co-rotating
components which often have larger amplitudes of rotation in the OASIS
data, corresponding to the nuclear disks visible on the HST
images. The effects of specific nuclear kinematics, related mean ages
of the components and possible different formation paths were
previously discussed in Paper VIII.

\begin{figure}
        \includegraphics[width=\columnwidth]{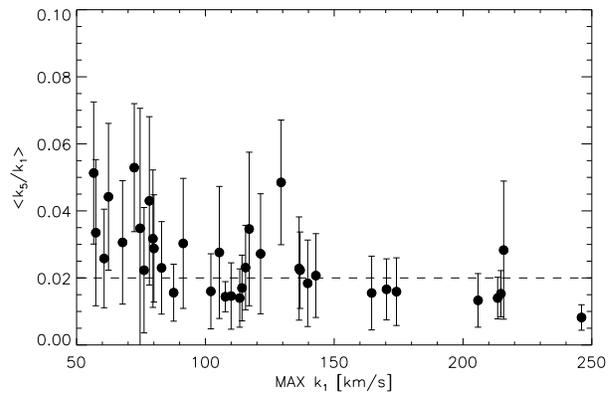}
  \caption{\label{f:k1_k51} A relation between the maximum amplitude
    of rotation and luminosity-weighted average of $k_5/k_1$ ratio,
    which measures departures from the cosine law for velocity
    profiles extracted along the best fitting ellipse. All points
    belong to fast rotators. }
\end{figure}

In addition to this grouping of the velocity maps, we can also
describe the kinematic components. Out of 15 SC galaxies, 8 are DR
(53\%), 2 are KT (13\%) and 4 are LV (27\%), while the remaining
galaxy (NGC2768) is a rather special case of a solid body rotator (see
below). The majority of MC galaxies have only two kinematic components
(30 or 91\%), but there are some with three kinematic components (3 or
9\%). Many of the components are similar in properties. If the inner
component ($C_1$) is DR then the second component ($C_2$) is also a
DR. If $C_1$ has a more complex kinematic structures (KDC, LV, KT),
$C_2$ or $C_3$ will in most cases still be a DR. Exceptions are found
in a few cases when $C_2$ is not rotating and can be described as LV.
In some cases components may show KT, but their $k_5/k_1$ ratio is
less than or equal to 0.02. Counting all galaxies that have at least
one component with $k_5/k_1\lesssim0.02$, the number of systems with
DR-like characteristics rises to 35 (73\%). In terms of E/S0
classification, this means that 74\% of Es and 92\% of S0s have
components with disk-like kinematics

Using the resolving power of the OASIS data we can go even further:
allowing for large error bars and allowing for considerable
uncertainty of component boundaries, virtually every galaxy that shows
rotation (including parts of KDCs) has at least one region with
$k_5/k_1\lesssim0.02$ (0.04 in OASIS). This can be seen on
Fig.~\ref{f:k1_k51} that shows luminosity weighted average of
$k_5/k_1$ ratio versus the maximum rotational amplitude $k_1$ for all
fast rotators measured on the SAURON data. The large uncertainties in
some cases reflect the multiple component nature of fast rotators,
because $k_5/k_1$ ratio rises in the transition region between
components \citep{2006MNRAS.366..787K}. Notably, both average values
and uncertainties rise as the maximum rotation velocity
decreases. This suggest a more complex structure (more components,
larger difference between components) in galaxies with lower amplitude
of rotation.

We estimated local kinematic misalignment for each kinematic
component. The results show that a total of 25 galaxies (52\% of the
sample) have some evidence of KM. It should, however, be kept in mind
that it is difficult to determine the sense of rotation for LV
components. Ignoring galaxies with a LV in the single component or in
the second component, we are left with 15 galaxies with kinematic
misalignments. Of these, 13 galaxies show misalignment in the first
component, and 7 in the second component. Moreover, only 5 galaxies
show misalignment in two components, as a global property within the
SAURON field of view.

As it can be seen from the kinemetric profiles there are special cases
for which kinemetry is not able to determine the characteristic
parameters robustly. This, in general, occurs when velocity drops
below $\sim15\kms$ (e.g. NGC4486, NGC4550), and it is not surprising
since these maps also often lack odd parity, an expected property of
the first moment of the LOSVD. As discussed in
\citet{2006MNRAS.366..787K}, another exception is the case of solid
body rotation for which the iso-contours are parallel with the
zero-velocity curve. In the SAURON sample this is seen in
NGC2768. Since the velocity iso-contours are parallel the axial-ratio
and the position angle are poorly defined and in practice strongly
influenced by noise in the data.  Determination of the best-fit
ellipse parameters for the solid body rotation is degenerate. From
these reasons this galaxy should also be considered with care when
comparing with kinemetry results for other objects.

\subsection{Fast vs Slow Rotators}
\label{ss:SRFR}

As stated above, all galaxies with $k_5/k_1 > 0.1$ along a significant
part of the profile from SAURON data are classified as slow rotators
in Paper IX. To some extent this is expected since slow rotators in
general show very little rotation and kinemetry assumptions are not
satisfied. The origin of the noise in the maps of slow rotators, which
generates large $k_5/k_1$, is likely reflecting a special internal
structure, and, in principal, does not come from technical aspects of
the observations. Even velocity maps with low amplitude of rotation
could show regular spider diagrams (e.g. disks seen at very low
inclinations) observed at the same signal-to-noise. The one-to-one
relation between slow rotators and objects with large higher-order
harmonic terms is significant since the slow/fast rotator
classification is based on both velocity and velocity dispersion maps
and reflects the internal structure.

The OASIS data cover only a small fraction of the effective radius and
do not show this relationship. The slow rotators NGC3414, NGC3608,
NGC5813 and NGC5982 show considerable rotation because the KDC is
covering the full OASIS FoV, while the central regions of the fast
rotators NGC2768, NGC3032 and NGC3379 still have small amplitudes of
rotation and high (and noisy) $k_5/k_1$ ratios. 

The velocity maps of the 12 slow rotating galaxies in our sample can
be described either as LV or as KDC+LV. In that respect, 4 slow
rotators are SC systems (NGC4374, NGC4486, NGC4550 and NGC5846) which
do not show any detectable rotation (at SAURON resolution) and the
other 8 are MC systems where $C_1$ is a KDC and $C_2$ is a LV. In
between these two cases is NGC4552 with a rather constant rotation
velocity of $15\kms$, the boundary level for LV, and the $C_1$ between
a KDC and a large KT. The other three quarters of galaxies in the
SAURON sample are fast rotators. Only a third of them are described as
SC, but as shown above (Section~\ref{ss:subcomp}) all fast rotators
have components with kinematics that can be described as DR. Moreover,
in the case of some slow rotators with small, but not negligible
rotation in the centres (KDC), the higher resolution OASIS data were
able to ascertain that these components have near to DR properties
(within often large uncertainties)

\begin{figure}
        \includegraphics[width=\columnwidth, bb=40 40 409 1036]{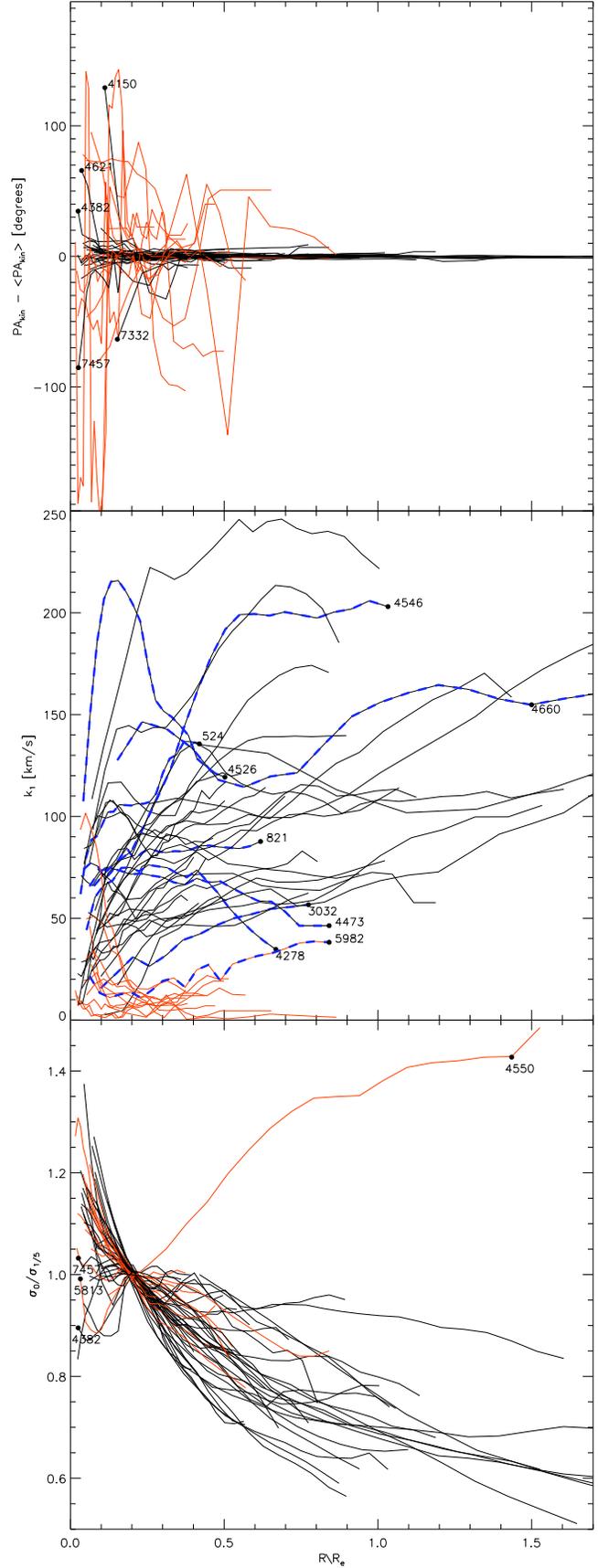}
  \caption{\label{f:radP} Radial profiles of (from top to bottom) \PAk
    - $\langle$\PAk$\rangle$, $k_1$ and $\sigma_0$ from the SAURON
    data. $\langle$\PAk$\rangle$ are luminosity weighted averages of
    \PAk~profiles.  $\sigma_0$ profiles are normalised at $R_e$/5. The
    profiles of slow and fast rotators are coloured in red and black,
    respectively. In the middle panel dashed blue lines are
    over-plotted to guide the eye for the cases with specific profiles
    as mentioned in text.}
\end{figure}

Figure~\ref{f:radP} has three panels highlighting most obvious
kinematic properties of slow and fast rotators. The top panel shows
radial variations of the kinematic position angles, \PAk, which are
present in various forms, ranging from minor twists in the nuclei,
through abrupt jumps at the end of KDCs, to almost random changes with
radius. However, only \PAk~of slow rotators are characterised by
strong and rapid changes. Fast rotators show remarkably constant
\PAk. If twists are present in \PAk of fast rotators, they are small
in amplitude ($\lesssim30\degr$) and confined to the nuclear region in
shapes of physically small KDCs (NGC4150, NGC4382, NGC4621, NGC7332,
NGC7457). Slow rotators on the other hand show a much greater
amplitude in change of \PAk.

It should be stressed again that determination of \PAk~for slow
rotators is much more ill defined than for fast rotators in the sense
that if there is no rotation, there is also no orientation of
rotation. The abrupt changes in \PAk~are the consequence of this in
some cases (NGC4374, NGC4486, NGC4550, NGC5846), and while one
could debate the robustness of measured \PAk, one should acknowledge
the different nature of these systems from objects with a constant
\PAk.

The difference between the slow and fast rotators is most visible in
the amplitude of rotation.  The middle panel of Figure~\ref{f:radP},
shows the radial profiles of $k_1$ kinemetric terms for 48 early-type
galaxies. Most of the profiles cover up to 1 $R_e$ in radius. Slow
rotating galaxies can show a non-zero amplitude of rotation in the
centres (KDCs), but the amplitude is, in general, not very high and
towards the edge of the map it is mostly negligible.  The only
exception is NGC5982, which approaches two fast rotators with the
slowest rotation in the outer regions (NGC4278 and NGC4473).

Another characteristic of this plot is the variety of profiles. They
include: monotonically rising profiles (e.g. NGC3032), profiles with
an initial slow rise which turns to a more rapid one (e.g. NGC524), a
rapid rise to a maximum followed by a plateau (e.g. NGC4546), rise to
a maximum followed by a decrease (e.g. NGC4526), double hump profiles
(e.g. NGC4660), flat profiles (e.g. NGC821) and, in slow rotators,
profiles showing a decrease below our detection limit. Keeping in mind
that the SAURON sample is not a complete sample, among fast rotators
there are 17 (47\%) with increasing profile at the edge of the SAURON
map, 9 (25\%) have flat profiles, 5 (14\%) decreasing profiles and 4
galaxies have intermediate (difficult to classify) profiles. Among
slow rotators there are 3 (25\%) galaxies with increasing profiles at
the edge of the map, the rest being flat and below the detection
limit.

These statistics are influenced by the size of velocity maps and the
coverage of kinematics components by kinemetric ellipses. Clearly,
larger scale observations would detect the end of rise in amplitude in
galaxies that are now observed to have increasing $k_1$. Similarly, it
is possible that a decrease in $k_1$ could be followed by an
additional increase or a flat profile at large radii. Still, there are
two general conclusions for fast rotators: they mostly show increasing
velocity profiles at $1 R_e$, where the range of maximum velocity
amplitude spans 200 $\kms$. On the other hand, slow rotators have
velocity amplitude mostly less than $20 \kms$ at $1 R_e$.

The bottom panel of Fig.~\ref{f:radP} shows radial velocity dispersion
profiles, $\sigma_0$, extracted along the isophotes from the velocity
dispersion maps as $a_0$ harmonic terms. All profiles are normalised
to their value at $R_e/5$. This highlights the similar general shape
of the $\sigma_0$ radial profiles. The only outlier is NGC4550 with an
$\sigma_0$ profile which increases with radius. Most of the other
profiles, while different in detail, show a general trend of
increasing $\sigma_0$ towards the centre and also have a similar
shape. A few profiles are consistent with being flat
($\Delta\sigma_0$/$\Delta R \lesssim 30 \kms$) over the whole profile
(visible only in fast rotators such as NGC7457).
 
If there are any real differences between $\sigma_0$ profiles, they
are apparent for radii smaller than $R_e/5$. There are a few
exceptions to the general trend: {\it i)} profiles with a decrease of
more than 5\% in the normalised $\sigma_0$ within $R_e/5$
(e.g. NGC4382), {\it ii)} profiles that are flat to within 5\% inside
the $R_e/5$ (e.g. NGC7457), and {\it iii)} profiles with a central
rise followed by a drop and consecutive rise forming a profile with
double maxima (e.g. NGC5813). These cases occur mostly in fast
rotators, with a few exceptions in slow rotators.

These central plateaus and drops are interesting, because classical
theoretical work predicts that, for constant mass-to-light ratio,
$r^{1/n}$ light-profiles have velocity dispersion minima in the
centres of galaxies \citep{1980MNRAS.190..873B, 1997A&A...321..724C},
unless they contain central black holes
\citep[e.g.][]{1998ApJ...498..625M}. The central $\sigma$-drops
evidently do not occur frequently in real early-type galaxies, but
are, perhaps marginally more common to fast rotators. About 10\% of
the SAURON early-types exhibit the central drop (but additional 20\%
have flat central profiles), which is much less compared to 46\% among
the Sa bulges, also observed with SAURON data
\citep{2006MNRAS.369..529F}.

\begin{figure*}
        \includegraphics[width=\textwidth, bb=50 40 870
          440]{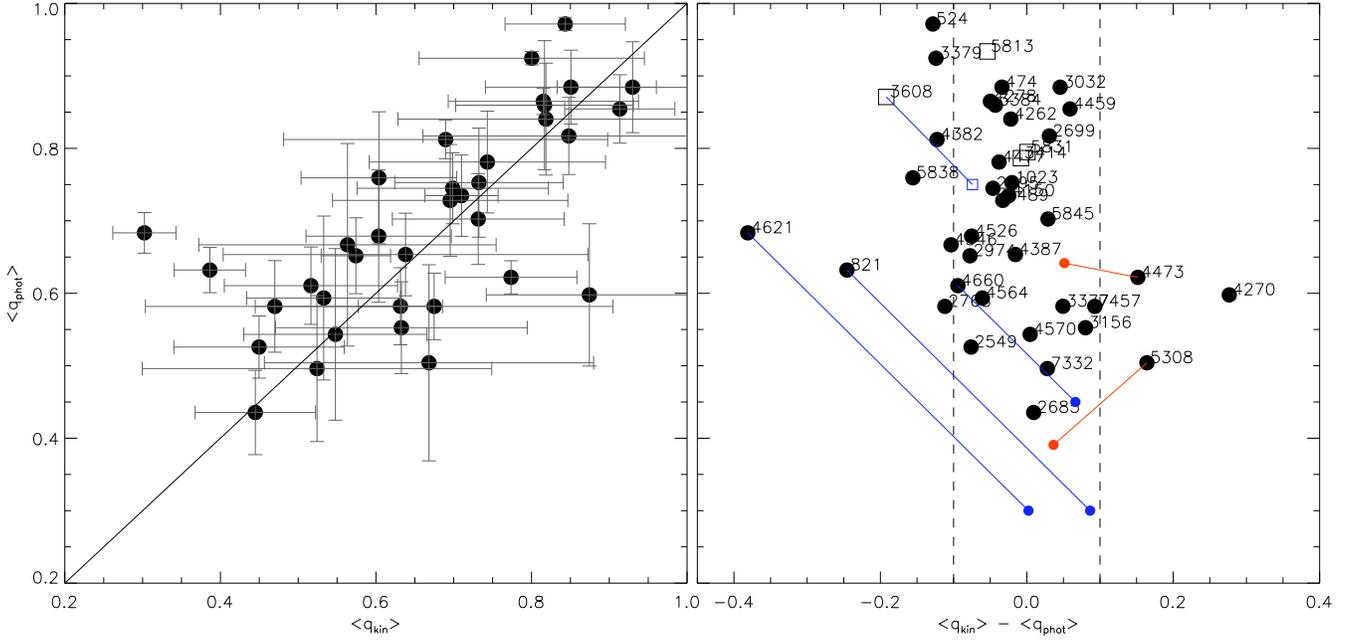}
  \caption{\label{f:qkvqp} {\bf Left:} Relation between
    luminosity-weighted kinematic and photometric axial ratios,
    $\langle$\qk$\rangle$ and $\langle$\qp$\rangle$, respectively, for
    fast rotating galaxies in the SAURON survey measured beyond
    $5\arcsec$ to avoid seeing effects. The black line is the 1:1
    relation. The error bars describe the radial variations of \qp~and
    \qk profiles. {\bf Right:} Difference between luminosity-weighted
    kinematic and photometric axial ratios plotted against the
    photometric axial ratio. Dashed vertical lines represent a typical
    variation of \qk~ profiles, which is the dominant uncertainty
    factor for comparison with \qp. Solid symbols are fast rotating
    galaxies, while open squares are KDC components of slow
    rotators. The red lines between symbols link $\langle$\qk$\rangle$
    with $\langle$\qp$\rangle$ measured on inner (NGC4473) and outer
    (NGC5308) kinematic components. The blue lines link
    $\langle$\qk$\rangle$ and axial ratio of the MGE models.}
\end{figure*}

\subsection{Distribution of axial ratios}
\label{ss:flat}

We now compare average photometric and kinematic axial ratios of
SAURON galaxies. The axial ratio of a velocity map is related to the
opening angle of the iso-velocity contours. In other words, the
pinching of the contours in a spider diagram is related to the axial
ratio of the best fitting ellipse given by kinemetry. As the kinematic
axial ratio of slow rotators is an ill defined quantity (set to 1),
in the rest of this section we focus on the average axial ratios of
the fast rotators.

In Fig.~\ref{f:qkvqp} we compare values of $\langle$\qk$\rangle$ and
$\langle$\qp$\rangle$ for fast rotating galaxies. 
Since the typical seeing for SAURON data ranges up to $2\farcs5$, we
exclude the inner $5\arcsec$ of the \qk~profiles from our derivation
of the luminosity weighted average values (Appendix~\ref{s:psf}). The
left hand panel shows a one-to-one correlation between the two
quantities, although the scatter and uncertainties are large. The
right hand panel shows more clearly the amount of scatter in these
relations, as measured by the difference $\langle$\qk$\rangle$ -
$\langle$\qp$\rangle$.  The typical variation of the measured average
$\Delta$\qk~is 0.1, as shown with vertical guidelines on the right
panel. Outside this region there are about dozen galaxies. A few of
these have $\langle$\qk$\rangle$ larger than $\langle$\qp$\rangle$;
their photometric axial ratio is flatter than the kinematic, while the
majority of outliers have the kinematic axial ratio flatter than the
photometric. Let us consider in more detail only objects at
significant distances from the vertical lines
(i.e. $\langle$qk$\rangle - \langle$qp$\rangle > 0.15$): NGC821,
NGC4270, NGC4473, NGC4621, NGC5308 and NGC5838.

If we look at the \qk~profiles of the three galaxies
(Fig.~\ref{f:vel_prof}) with $\langle$\qk$\rangle >
\langle$\qp$\rangle$, we can see that some of their kinematic
sub-components have axial ratios very similar to the local photometric
axial ratios. Notably, in the case of NGC5308 this is the outer
component, especially at radii near to the edge of the SAURON FoV. The
middle range, where the differences are the largest, is also the
region of the transition between the two kinematic components. The
mixing of the components changes the measured $\langle$\qk$\rangle$,
but it should be also noted that the $\langle$\qp$\rangle$ varies over
the whole map, becoming flatter and similar to $\langle$\qk$\rangle$
with radius. This is not the case in NGC4473. Here the photometric
axial ratio remains constant, but the kinematic axial ratio changes at
larger radii. Again this change occurs in the transition region
between the two kinematic components. Detailed dynamical modelling of
this galaxy shows that it is made of two counter-rotating stellar
components of unequal mass. This object is physically similar to
NGC4550, where the main difference is in the mass fraction of the two
components (Paper X). In the case of NGC4270 it is the \qp~that
steadily changes with radius, while \qk~has mostly high values, but
shows abrupt changes between the points. These are related to
sometimes rather high values of $k_5/k_1$, which also changes abruptly
between the adjacent points, a behaviour originating from the noisy
transition region between two kinematic components. Since it is hard
to disentangle the noise from the genuine physical signal in \qk~
measurement, and given the boxy appearance on the large scales, we
note that this object could be a true and unusual outlier from the
relation between \qp~and \qk.

Galaxies with $\langle$\qk$\rangle<\langle$\qp$\rangle$ are either
single component (NGC821 and NGC4621, if we exclude the CRC in NGC4621
of $\sim 4\arcsec$ in size) or multi component (NGC5838). The HST
image of NGC5838 has a prominent nuclear dust disk with the axial
ratio between 0.3 -- 0.4, constraining its inclination to about
$70\degr$. It is possible that the velocity map is dominated by the
presence of an associated stellar disk embedded in the galaxy
body. NGC821 and NGC4621 were parametrised via the Multi-Gaussian
Expansion \citep{1994A&A...285..723E, 2002MNRAS.333..400C} in Paper
IV. Both models required very flat Gaussians to reconstruct the light
distribution (in both cases, the smallest axial ratios of the
Gaussians of the MGE models were 0.3); these Gaussians are tracing
disks embedded in spheroids. If we compare the $\langle$\qk$\rangle$
with the flattest MGE Gaussians, both galaxies move well within the
two vertical lines on right hand panel of Fig.~\ref{f:qkvqp} (blue
lines).

As a matter of interest, on the same right hand plot of
Fig.~\ref{f:qkvqp} we overplotted axial ratios for the big KDCs in the
sample: NGC3414, NGC3608, NGC5813 and NGC5831. The shape of their
kinematics is similar with the distribution of light, except in the
case of NGC3608. Its KDC has flatter kinematics than the light, but
the flattest MGE Gaussian is comparable with the kinematic axial
ratio. This suggest that even some of the sub-components of the slow
rotating galaxies have similar properties like fast rotating galaxies.

Having in mind that \qk~profiles can vary significantly with the
radius, and the average values could be contaminated by the
contribution of the transition regions between the components, we
conclude that there is a near one-to-one correlation between average
kinematic and photometric axial ratios in fast rotating galaxies, with
a number of objects having $\langle$\qk$\rangle <
\langle$\qp$\rangle$, and, hence, having disk-like components more
visible in their kinematics than in photometry.

\begin{figure*}
        \includegraphics[width=\textwidth,bb=40 35 850
          440]{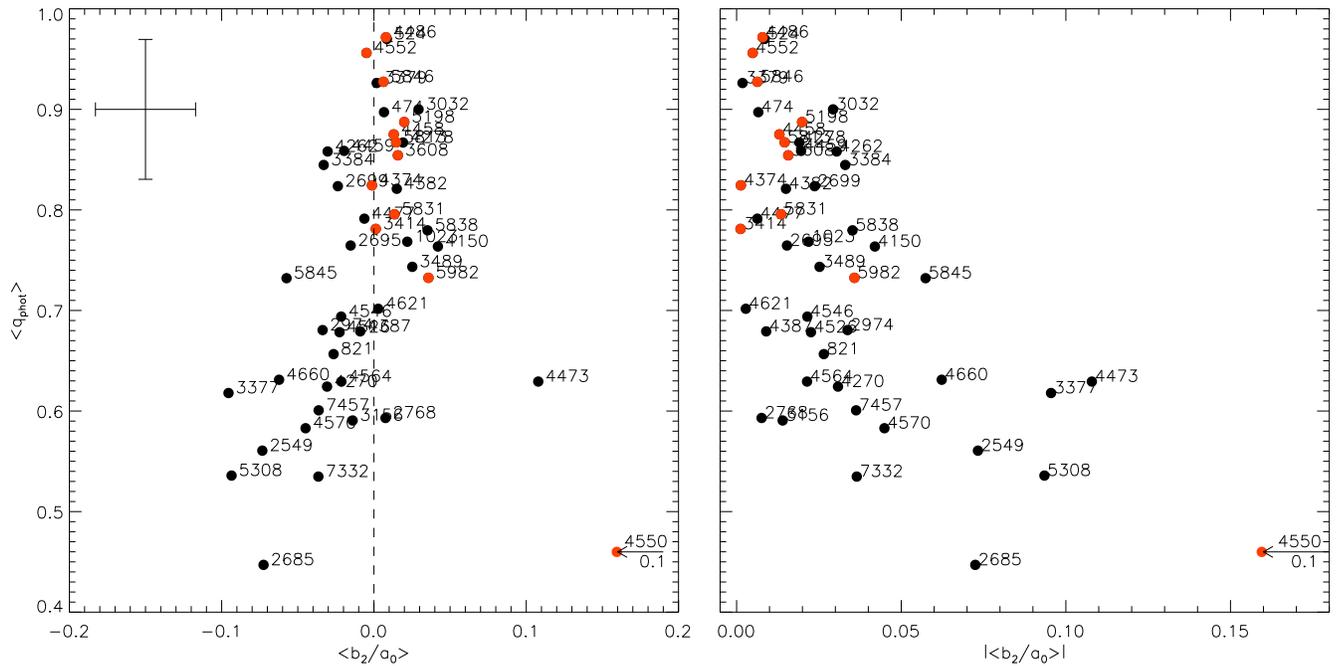}
  \caption{\label{f:b2a0} {\bf Left:} Relation between isophotal axial
    ratio and the luminosity weighted average normalised second term
    ($\langle b_2/a_0 \rangle _\sigma$) of the Fourier decomposition
    of the velocity dispersion profiles extracted along the
    isophotes. Red symbols are slow rotators. An average uncertainty,
    describing the radial variation of $b_2/a_0$ profiles, is shown in
    the upper left corner. Large negative $\langle b_2/a_0 \rangle
    _\sigma$ values are typical for iso-$\sigma$ contours that are
    rounder than the isophotes, while large positive $\langle b_2/a_0
    \rangle _\sigma$ values are typical for flatter iso-$\sigma$
    contours than the isophotes.  {\bf Right:} same as left, but now
    absolute values of $\langle b_2/a_0 \rangle _\sigma$ are
    plotted. In both images, NGC4550 was shifted to the left for 0.1
    for presentation purposes, as shown by the arrows.}
\end{figure*}

\subsection{Shape differences between velocity dispersion and surface brightness maps}
\label{ss:shape}

As shown in Section~\ref{ss:prof}, the isophotes are not necessary a
good representation of contours of constant velocity dispersion and
the deviations are visible in the second cosine term ($b_2$) of the
harmonic decomposition of the velocity dispersion profiles extracted
along the isophotes. On Fig.~\ref{f:b2a0} we quantify the differences
between isophotes and iso-$\sigma$ contours by plotting the normalised
luminosity-weighted second term ($\langle b_2/a_0 \rangle _{\sigma}$)
extracted along the isophotes of SAURON galaxies. Slow rotators are
shown in red.

Focusing on the left hand panel, it is clear that galaxies in the
sample span a large range of $\langle b_2/a_0 \rangle _{\sigma}$, both
positive and negative. If we exclude NGC4473 and NGC4550, there is a
tail of galaxies with negative values up to $\sim-0.1$. Also, it seems
that there is a trend of high negative $\langle b_2/a_0 \rangle
_{\sigma}$ in flat galaxies: as galaxies become rounder, $\langle
b_2/a_0 \rangle _{\sigma}$ tends to zero; above
$\langle$\qp$\rangle=0.8$ galaxies have small absolute value of
$\langle b_2/a_0\rangle _{\sigma}$. Slow rotators are relatively round
systems (Paper X) with \qp $> 0.7$ (excluding the special case of
NGC4550). Their $\langle b_2/a_0\rangle _{\sigma}$ values are small
and mostly positive. Typical measurement error of $\langle b_2/a_0
\rangle _\sigma$ is $\sim 0.03$, and hence, the slow rotators are
consistent with having velocity dispersion maps very similar to the
the distribution of light, with possibly marginally flatter
iso-$\sigma$ contours.

Before we turn to the right panel on Fig.~\ref{f:b2a0}, let us go back
to Fig.~\ref{f:diff} and the example of NGC2549 and NGC4473. Their
stellar velocity dispersion maps have different shapes from the
distributions of light, but they have similar absolute values of
$\langle b_2/a_0 \rangle _{\sigma}$; they fall on the opposite sides
of Fig.~\ref{f:b2a0}. There is, however, evidence that in both cases
the deviations from the photometry have similar physical
origin. NGC2549 shows clear photometric and kinematic evidence for a
stellar disk viewed at a high inclination (Fig.~\ref{f:vel_prof}). The
velocity dispersion map has a very specific 'bow-tie' shape in the
central 10\arcsec, while outside its amplitude is low everywhere. The
'bow-tie' shape can be explained with a light-dominating dynamically
cold stellar disk which decreases the observed velocity
dispersion. Outside of the disk, the bulge dominates and the observed
velocity dispersion is again higher. It should be noted that the view
of the galaxy at a high inclination is crucial for this effect to be
seen.

\begin{figure}
  \epsfxsize=0.79\hsize \epsfbox[50 45 427 440]{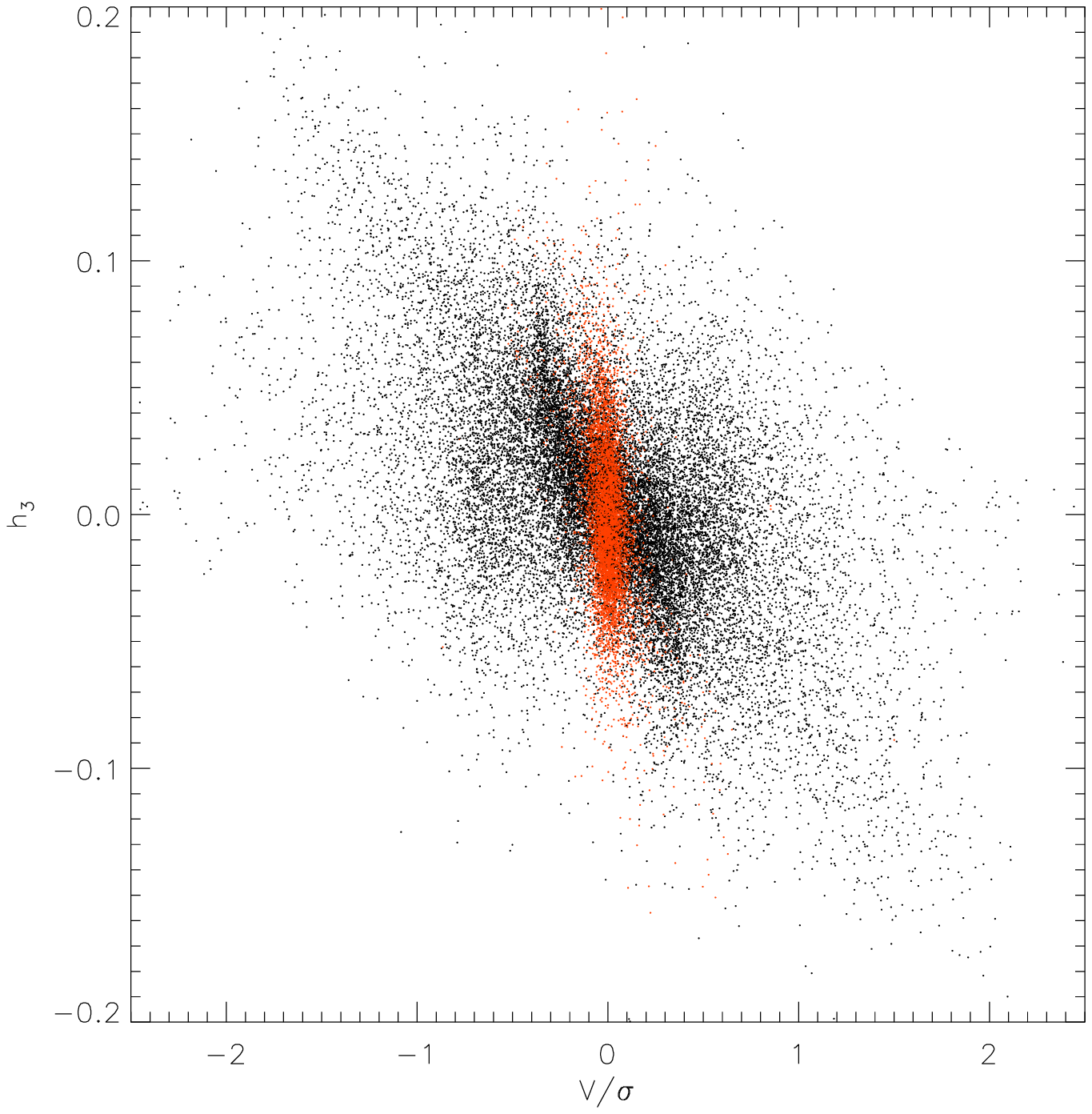}
\hspace{0.5cm}
  \epsfxsize=0.79\hsize \epsfbox[50 45 427 440]{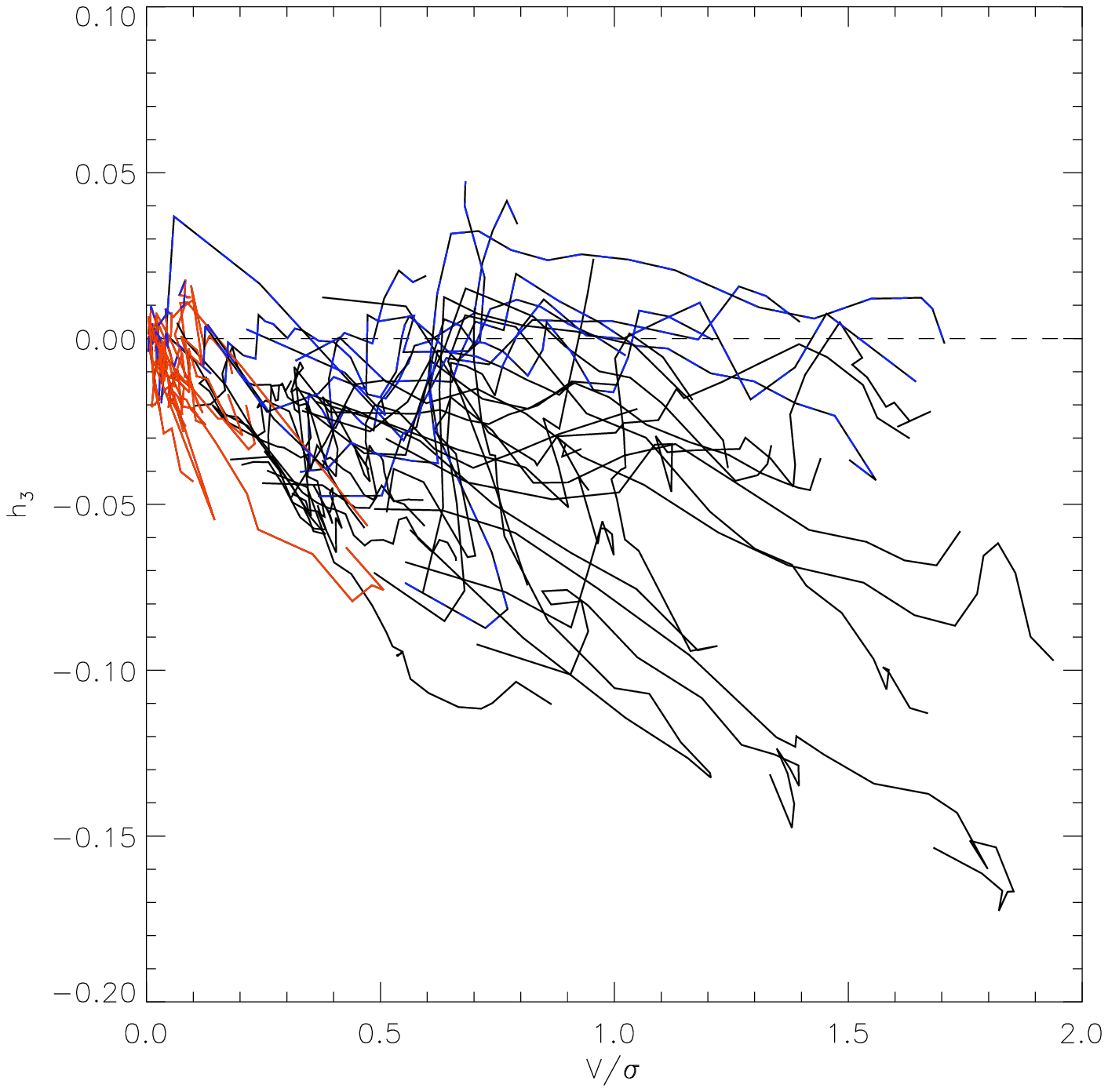}
\hspace{0.5cm}
  \epsfxsize=0.79\hsize \epsfbox[50 45 427 440]{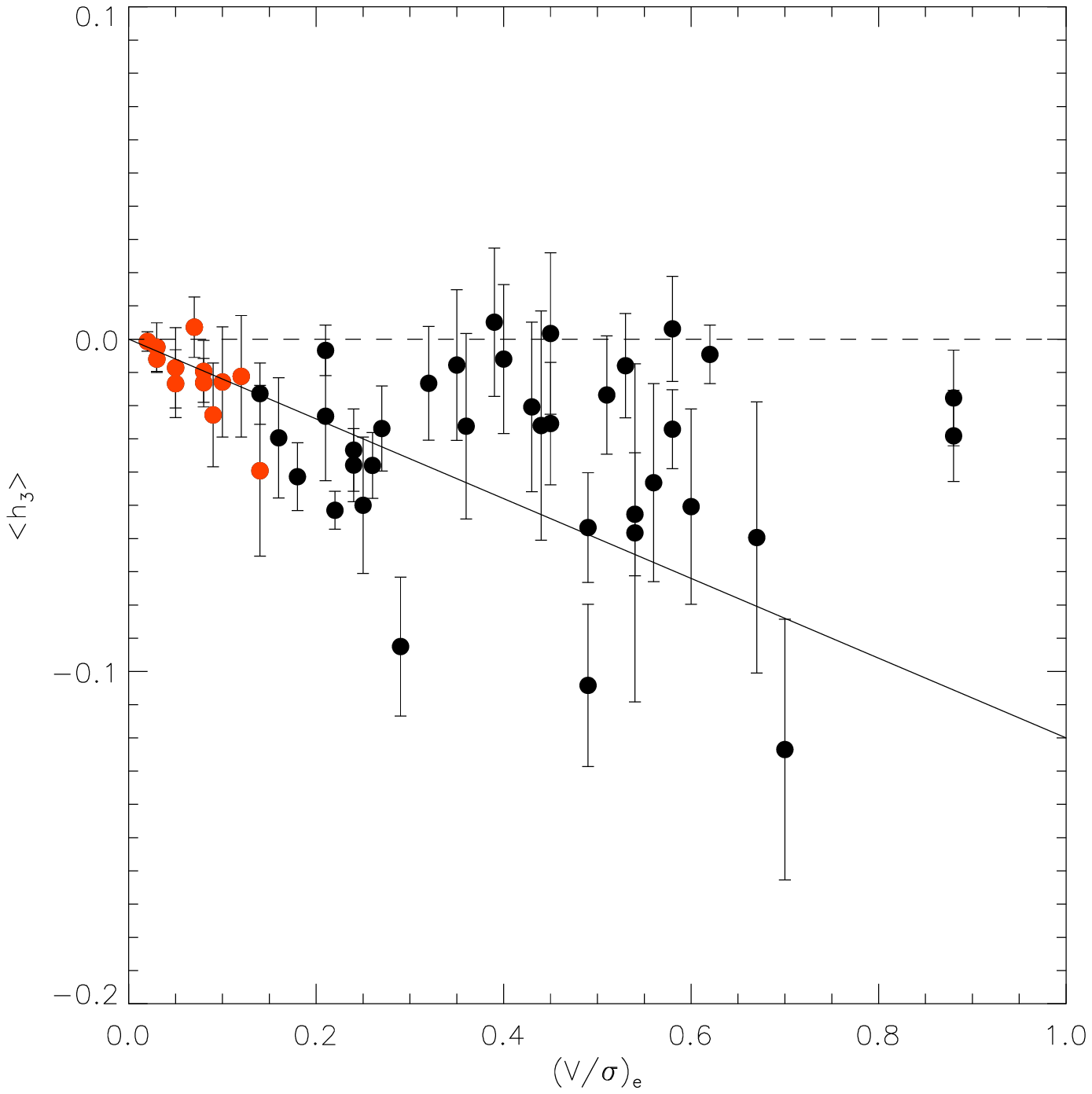}
  \caption{{\bf Top:} Local $h_3 - V/\sigma$ relation for all data
    points of the 48 SAURON galaxies with $h_3$ error less than
    0.2. {\bf Middle:} Profiles of $h_3 - V/\sigma$ relation averaged
    on ellipses. On all maps ($V$, $\sigma$ and $h_3$) profiles were
    extracted along the best fitting ellipses to the velocity
    maps. Typical error bars are of the order of 0.01 or smaller, due
    to averaging along the best fitting ellipse. {\bf Bottom:}
    Relation between the luminosity averaged $h_3$ and $(V/\sigma)_e$
    ($V/\sigma$ within 1 $R_e$ from Paper X). The error bars describe
    the radial variation of $h_3$ profiles. The solid line is the
    relation from \citet{1994MNRAS.269..785B}. On all panels, slow
    rotators are shown in red.}
\label{f:h3_prof}
\end{figure}
NGC4473 was discussed in the previous section where we stated that it
actually contains two sub-components. One of them is flat and
counter-rotates with respect to the main galaxy body. This can explain
the rise of the velocity dispersion along the major axis, specifically
at larger radii where the contribution of the flat component becomes
similar to the main body. This is also visible in the mean velocity
map, where beyond 10\arcsec~the velocity starts dropping. The only
significant difference between these two examples is in the sense of
rotation of the flat sub-components. Co-rotating disks or flattened
components, viewed at favourable angles will likely show negative
$\langle b_2/a_0 \rangle_\sigma$, while counter-rotating components
will contribute to positive $\langle b_2/a_0 \rangle _\sigma$. In
other words, co-rotation increases, while counter-rotation decreases
the flattening of the iso-$\sigma$ contours.

A similar case is NGC4550, the most extreme outlier, which also has
two counter-rotating disks. NGC4150 and NGC3032 can be put in the same
group: the OASIS measurements resolve their CRC components. It is
interesting to note that the strongest $\sigma$-drop galaxies (NGC2768
and NGC4382) are both found on the positive side of the left hand
panel of Fig.~\ref{f:b2a0}, admittedly with small $\langle b_2/a_0
\rangle _\sigma$ ($<0.02$), indicating that the shapes of the velocity
dispersion maps of $\sigma$-drop objects are somewhat different from
the shapes of galaxies with central plateaus in velocity dispersion
profiles or bow-tie shape of velocity dispersion maps. The shapes of
$\sigma$-drops are more similar to the isophotes than the shape of
iso-$\sigma$ contours in the objects with strong near-to edge-on disks
such as NGC2549 or NGC3377. If $\sigma$-drops, bow-ties and central
plateaus are all caused by disks, than the difference between them
might originate from the orientation of the galaxy, where bow-ties are
seen at edge-on and $\sigma$-drops at face on orientations.

Given the uncertainties, it is hard to argue either way about other
galaxies with slightly positive $\langle b_2/a_0 \rangle _\sigma$
values. Other effects, such as the actual size of the components or
the presence of dust, may play a role on that level. This should be
noted, but we, however, continue by suggesting that the main
contributor to the shape difference between the zeroth and the second
moment of the LOSVD in early-type galaxies are the embedded disks or
flattened fast rotating components seen at different inclination. In
this respect, on the right hand panel of Fig.~\ref{f:b2a0} we plot the
absolute value of $\langle b_2/a_0\rangle _\sigma$. Here a trend is
clear: from round slow rotators with generally small absolute values
to flat fast rotating galaxies with increasing $|\langle b_2/a_0
\rangle _\sigma|$. This sequence is the one of increasing contribution
of the ordered rotation towards the total energy budget, but it is,
unfortunately, dependent on the viewing angles (NGC524 with the
face-on disk in the top left corner of the diagram) and on the
luminosity-weighted contribution of the flat component to the total
kinematics (CRC in NGC4621 is not detected).

\subsection{Higher order moments of the LOSVD}
\label{ss:high}

The final part of our analysis of the kinematic moments of the LOSVD
is devoted to maps of $h_3$ and $h_4$, the Gauss-Hermite moments which
measure deviations of the LOSVD from a pure Gaussian. They were
introduced because LOSVD profiles are rather non-Gaussian,
specifically the contribution of $h_3$, the skewness, along the
major-axis of galaxies was found to have large amplitudes
\citep[e.g.][]{1994MNRAS.268..521V}. In a diagram of local $h_3 -
V/\sigma$ relation, \citet{1994MNRAS.269..785B} plotted the major axis
points of their objects which could be separated in two distinct
groups made of galaxies with disky isophotes and non-disky
isophotes. Both groups showed anti-correlation between $h_3$ and
$V/\sigma$, where the points with small $V/\sigma$ values in the
centre of the diagram had a steeper slope. It was recognised that the
disky early-type galaxies cover a large range in $V/\sigma$ and are
responsible for tails in upper left and lower right corner of the
diagram, while non-disky early-types with small $V/\sigma$ values
showed a somewhat smaller range in $h_3$ values.

On the top panel of Fig.~\ref{f:h3_prof} we plot the local $h_3 -
V/\sigma$ relation for all data points of 48 SAURON galaxies. Red
points belong to Voronoi bins of the slow rotators. This plot is
somewhat different from previous findings. Paper X showed that slow
rotators have small global $V/\sigma$ and this is also reflected in
bin by bin values. The black points represent the Voronoi bins of fast
rotators, and their distribution is different from what was found
before for disky ellipticals. The shape of the distribution of black
points can be described as a superposition of two components: one
which is anti-correlated with $V/\sigma$ and makes distinct tails in
upper left and lower right corner of the diagram (large positive and
negative $V/\sigma$ values), and the other that is both consistent
with $h_3\sim0$, and shows positive correlation at intermediate
$V/\sigma$.

Existence of $h_3 \sim 0$ distribution of points could be explained by
the fact that, while \citet{1994MNRAS.269..785B} plot only the points
along the major axis, we plot them all, and the horizontal
distribution and correlating tails are the consequence of the minor
axis contamination. Using kinemetry we extracted profiles of the
dominant harmonic terms from $V$, $\sigma$, and $h_3$ maps along the
same best-fitting ellipses to the velocity maps\footnote{Note that
  here we re-run kinemetry on $\sigma$ maps along the best fitting
  ellipses for velocity maps, except in the case of the slow rotators
  when we used circles as explained in Section~\ref{ss:prof}. These
  profiles are somewhat different from the profiles presented in
  Section~\ref{ss:shape} which were obtained along the isophotes.}. We
plot the $h_3 - V/\sigma$ profiles on the middle panel of
Fig.~\ref{f:h3_prof}. These profiles are basically major-axis
representation of two-dimensional maps and are more comparable with
previous major axis data. The slow rotators (in red) have the smallest
$V/\sigma$ values and generally small amplitudes of $h_3$. The two
trends for fast rotators from the left-hand panel are still
visible. Fast rotators cover a range of $V/\sigma$ values, but some
have large and some small $h_3$ amplitudes for a large $V/\sigma$
values. In some specific cases there is a suggestion that $h_3$
changes the sign becoming positive and correlating with $V/\sigma$
(e.g. NGC5308) at larger radii. 

A confirmation of this can also be seen on the bottom panel of
Fig.~\ref{f:h3_prof}. Here we plot the relation between luminosity
averaged $h_3$ and $(V/\sigma)_e$ obtained within $1 R_e$, as
advocated by \citet[][values taken from Paper
X]{2005MNRAS.363..937B}. The shown error bars are not statitcial
uncertainties but describe the radial variation of $h_3$
profiles. This plot can be compared to Fig.14b of
\citet{1994MNRAS.269..785B}, which also shows galaxies with $\langle
h_3 \rangle\sim 0 $ for intermediate $V/\sigma_m$, while their
empirical fitting relation (the solid line) describes the general
trend in our data.

\begin{figure*}
  \epsfxsize=0.495\hsize \epsfbox{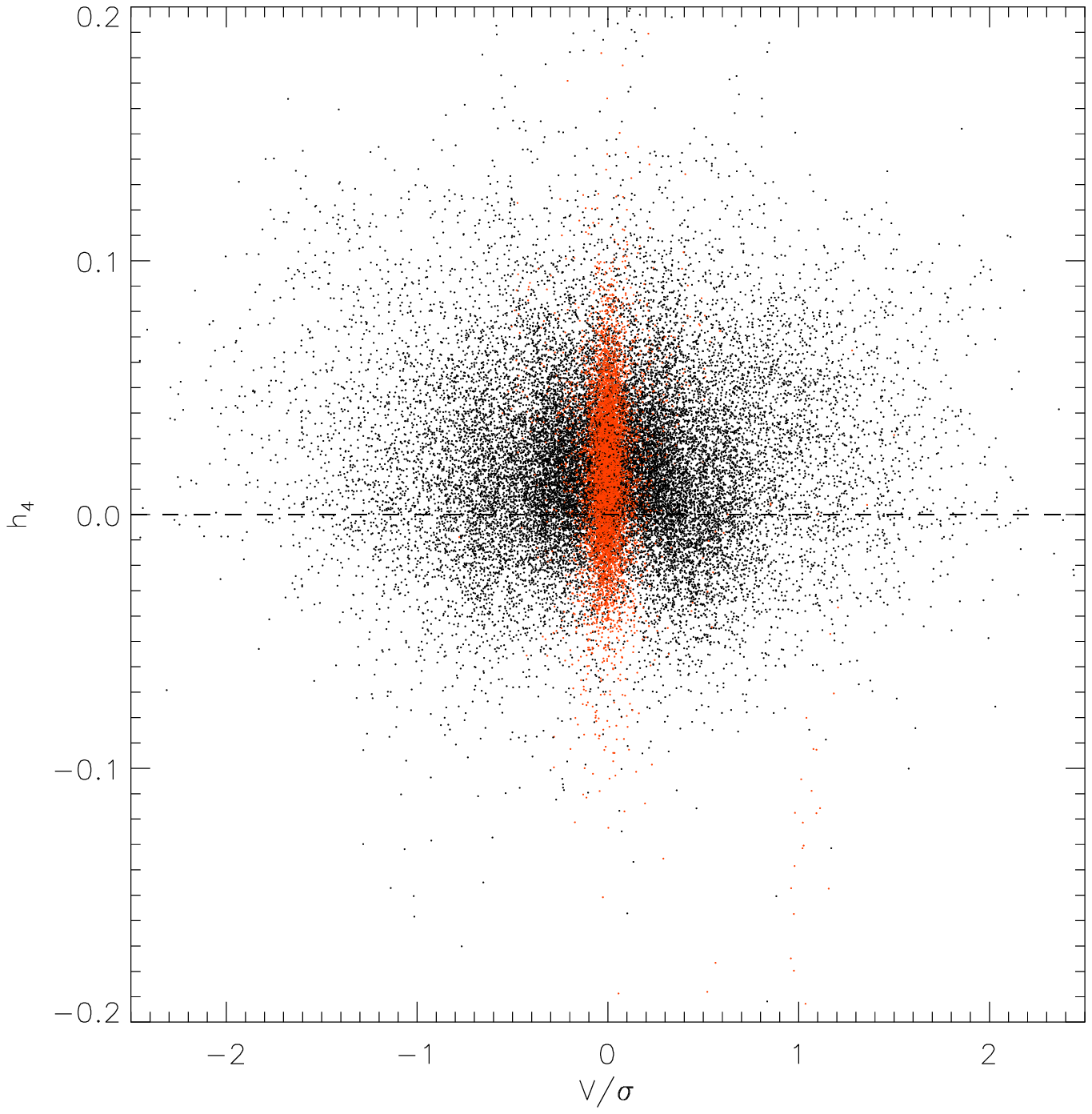}
  \epsfxsize=0.495\hsize \epsfbox{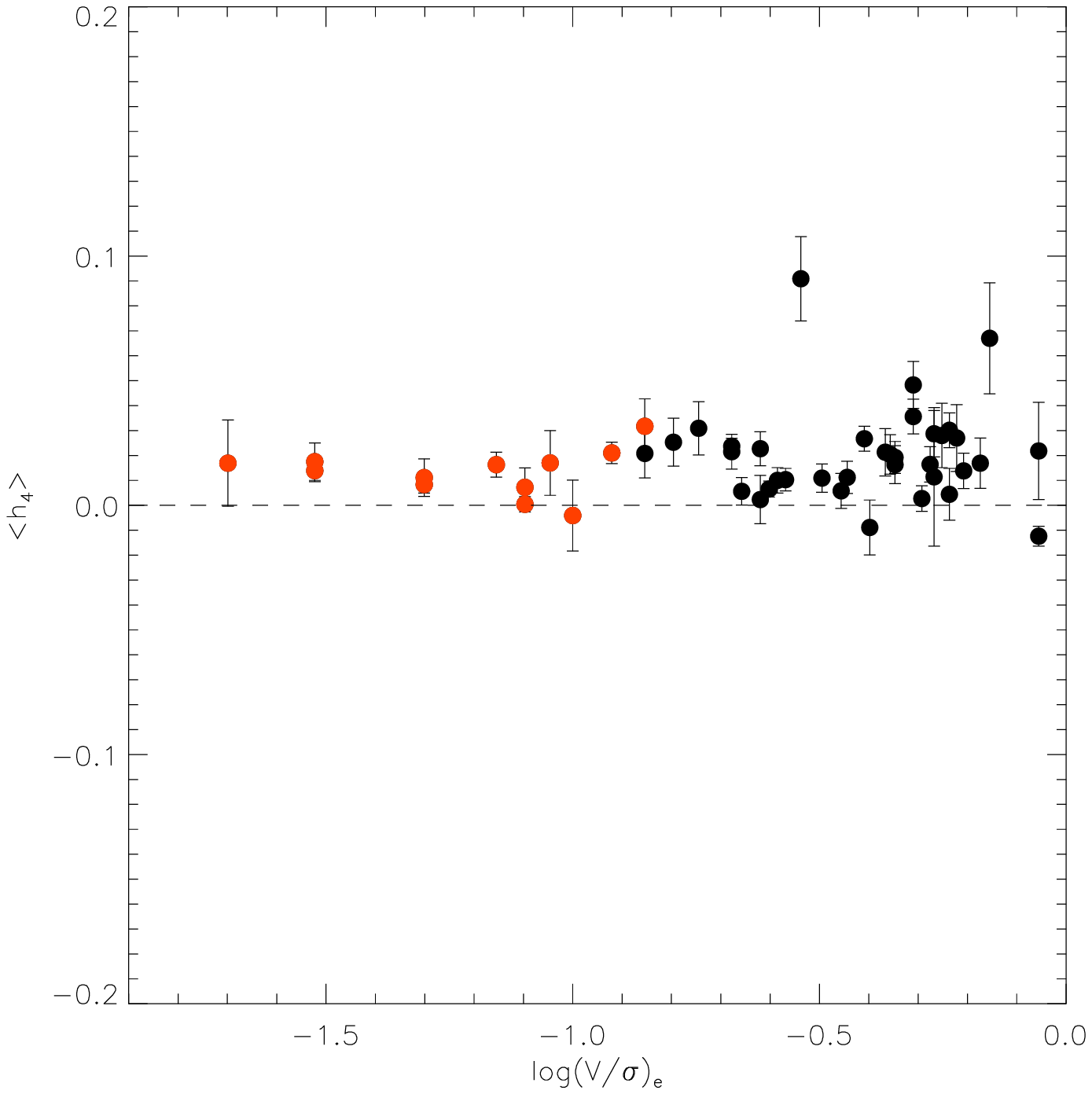}
  \caption{\label{f:h4} {\bf Right:} Local correlation between $h_4$
    and $V/\sigma$ for all data points of the 48 SAURON galaxies with
    $h_4$ error less than 0.2.  {\bf Right:} Correlation between
    luminosity weighted values average $\langle h_4 \rangle$ and
    $(V/\sigma)_e$ (Paper X). Red symbols are slow rotators in both
    panels. The error bars describe the radial variation of $h_4$
    profiles.}
\end{figure*}

The last kinematic moment to be analysed is given by the $h_4$
Gauss-Hermite coefficient. This moment describes the symmetric
departures of the LOSVD from a Gaussian. It should be, however, kept
in mind that it is very difficult to measure $h_4$ robustly, since it
strongly depends on the effect of template mismatch. Also an
inaccurate removal of the continuum will cause spurious $h_4$ values.

On the left hand panel of Fig.~\ref{f:h4} we plot the local $h_4 -
V/\sigma$ relation. Slow rotators with small $V/\sigma$ values,
dominate the central region. They cover a range of positive $h_4$
values and somewhat extend below zero. Fast rotators fill in a cloud
around slow rotators, equally filling negative and positive $V/\sigma$
part of the plot. There is a suggestion for a larger spread in
$V/\sigma$ for positive $h_4$ values. Luminosity weighted average
$\langle h_4\rangle$ values extracted with kinemetry along the
isophotes are presented as a function of luminosity-weighted
$V/\sigma$ measured within 1 $R_e$ (Paper X) in right panel of
Fig.~\ref{f:h4}. As $(V/\sigma)_e$ increases, there is a marginal
trend of an increased spread in the observed $h_4$ values, and
galaxies with negative average values start to appear. This result is
similar to \citet{1994MNRAS.269..785B}, in the sense that negative
$\langle h_4\rangle$ appear for larger $(V/\sigma)_e$, but we do not
find as negative values of $\langle h_4\rangle$. The only slow rotator
with negative $\langle h_4 \rangle$, however, is the usual outlier,
NGC4550, the special case of two counter-rotating disks.

%
%
\section{Discussion}
\label{s:disc}

\subsection{Velocity maps of slow and fast rotators}
\label{ss:velmapSRFR}

Radial profiles of kinemetric coefficients show that early-type
galaxies are {\it (i)} multicomponent systems and {\it (ii)} in the
majority of the cases contain a kinematic equivalent of a disk-like
component. These statements are based on the empirical verification
that the assumption of kinemetry holds. The assumption is that the
azimuthal profiles, extracted from a velocity map along the best
fitting ellipses, can be described with a simple cosine
variation. Practically, this means that higher-order harmonic terms
are negligible and at our resolution we find that the deviation from
the pure cosine law is less than 2\%, for about 80\% of cases, at
least on a part of the map. The multiple components are visible in the
abrupt and localised changes of kinemetric coefficients. The central
regions often harbour separate kinematic components which co- or
counter-rotate with respect to the outer body.  In some cases there
are more than two components (e.g. NGC4382) or components are similar
in size and co-spatial (e.g. NGC4473). Although the axial ratio
profile in edge-on systems can change with the seeing effects (see
Appendix~\ref{s:psf}), and in some cases the seeing can alter the map
considerably, the kinematic sub-components are usually robust
features.

There are, however, early-type galaxies for which the deviations from
the cosine law exceed 10\% across a significant radial range. Such
objects are all classified as slow rotators in Paper IX. The breakdown
of the kinemetry assumptions is another evidence that these objects
are intrinsically different. Certainly, we might not be able to apply
kinemetry {\it a priori} in its odd version on velocity maps that do
not show odd parity (e.g. NGC4486), but in the case of slow rotators
that show a detectable level of rotation (e.g. NGC5982), the
kinemetric analysis clearly shows differences from the maps of fast
rotators (e.g. NGC4387).

There are three types of intrinsic structures that will show small
mean velocities: {\it (i)} face-on disks, {\it (ii)} two
counter-rotating equal in mass and co-spatial components and {\it
  (iii)} triaxial structures.

\noindent {\it (i) Face-on Disks.} Low inclination thin disks will
still have minimal deviations from the cosine law assumed in
kinemetry.  Although the amplitude of the rotation in such cases is
small, the disk-like rotation (negligible $k_5/k_1$ coefficients) is
expected at all projections except in the actual case of $i=0\degr$
with no rotation. Such an example in the sample is NGC524 ($i \sim
19\degr$). Based only on the velocity map, one could naively interpret
that NGC4486 is a thin disk at $i=0\degr$, given that it has \qk$~\sim
1$ and shows no rotation. However, other moments of the LOSVD strongly
rule out this geometry, while, in general, other slow-rotators are not
round enough to even be considered as such extreme cases.

\noindent {\it (ii) Counter-rotating components.} This case can be
outlined with NGC4550 and NGC4473. These galaxies are examples of
axisymmetric objects with two counter-rotating and co-spatial
components. The first one is classified as a slow rotator, while the
other one is a fast rotator. In both cases, however, it is the mass
fraction of the components that really decides what is observed. In
the case of NGC4550 the masses are nearly equal and the luminosity
weighted mean velocity is almost zero in the central region. NGC4550
is a product of a very specific formation process, but a clear example
of how a superposition of two fast rotating components can imitate a
slow rotator. In NGC4473, on the other hand, one component is more
massive and dominates the light in the central $10\arcsec$, where the
rotation is clearly disk like. Outside this region, where the
counter-rotating component starts to significantly contribute to the
total light, the shape of the velocity map changes, the amplitude of
the rotation drops and non-zero $k_5/k_1$ coefficients are necessary
to describe stellar motions.

\begin{figure*}
        \includegraphics[width=\textwidth, bb=45 40 862
          437]{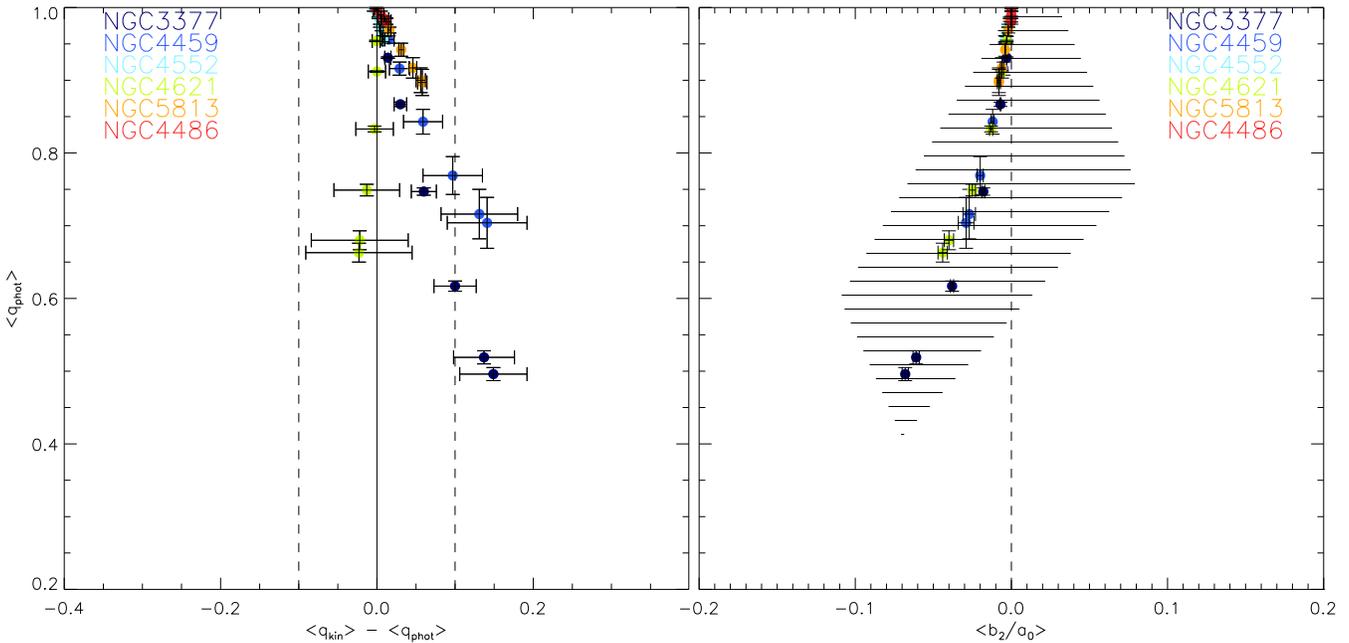}
        \caption{\label{f:jeans} {\bf Left:} Difference between
          luminosity weighted kinematic and photometric axial ratio,
          $\langle$\qk$\rangle$ and $\langle$\qp$\rangle$, for 5 Jeans
          models at $25\degr$, $35\degr$, $50\degr$, $65\degr$,
          $80\degr$ and $90\degr$ inclinations plotted against the
          luminosity weighted photometric axial ratio. The increase in
          inclination decreases $\langle$\qp$\rangle$ and moves points
          downwards. The vertical dashed lines are guidelines of the
          variation of \qk~profiles from the SAURON data as in
          Fig.~\ref{f:qkvqp}. Vertical solid line shows the location
          of zero difference between $\langle$\qp$\rangle$ and
          $\langle$\qk$\rangle$. {\bf Right:} Relation between
          photometric axial ratio and absolute $\langle b_2/a_0
          \rangle _\sigma$ for the same Jeans models as on the
          left. With increasing inclination the points move towards
          more negative $\langle b_2/a_0 \rangle _\sigma$ values. The
          hatched area represents the location of the data from
          Fig.~\ref{f:b2a0}}
\end{figure*}

\noindent {\it (iii) Triaxial objects.} Observationally, these objects
are marked by kinematic twists and kinematic misalignments, which are
not present in axisymmetric galaxies. A strong restriction to the
shape of the velocity maps of axisymmetric galaxies is that they
should not have a radial variation of \PAk~or a misalignment between
kinematic and photometric position angles. This, however, is not the
case for triaxial galaxies, where the change of position angle can be
influenced by a true change of the angular momentum vector, by the
orientation of the viewing angles or by the relative dominance of
different orbital families \citep{1991ApJ...383..112F,
  1991AJ....102..882S, 2008MNRAS.385..647V}. Paper X found that,
globally, kinematic misalignments are present only in slow rotating
galaxies. Locally, however, we find both kinematic misalignments and
twists in some fast rotating objects as well, but they are mostly
confined to central components or clearly related to bars
(e.g. NGC1023) or galaxies with shells (e.g. NGC474).

We do not find strong kinematic twists typical of the extreme cases of
maximum entropy models of triaxial galaxies projected at various
viewing angles (\citep{1991AJ....102..882S}; see also
\citet{1994MNRAS.271..924A}) for velocity maps of St\"ackel triaxial
models). Given that our sample is not representative of the luminosity
function of local early-type galaxies in the sense that it contains
too many massive galaxies, which are also more likely to show extreme
features on the velocity maps, it is remarkable that we find that only
a few velocity maps are similar to those predicted. Still, the
observed velocity maps are divers (e.g. the difference between the
maps of slow and fast rotators) and their complexity reflects the
difference in their internal structure.

The profiles of the relative change of the kinematic position angles
from the first panel of Fig.~\ref{f:radP} suggest that fast and slow
rotators have genuinely different intrinsic shapes, fast rotators
being mostly axisymmetric and slow rotators weakly triaxial. This is
also reflected in $k_5/k_1$ ratio. We suggest that the high values of
this ratio in slow rotators, which is in practice caused by the noise
in non-rotating velocity maps, has its origin in the internal orbital
make up of these galaxies. Weakly triaxial slow rotators contain box
orbits, and competing contributions of different tube orbit families,
as opposed to more axisymmetric fast rotators with short-axis tubes as
the only major orbit family \citep{1985MNRAS.216..273D}\footnote{We
  should keep in mind that two short-axis tube families with opposite
  angular momentum in certain cases can produce axisymmetric objects
  which appear as slow rotators, as mentioned above in the case of
  NGC4550.}.This suggestion is also supported by the analysis of the
orbital structure of collisionless merger remnants
\citep{2005MNRAS.360.1185J} as well as by the kinemetric analysis of
velocity maps of simulated binary disc merger remnants
\citep{2007MNRAS.376..997J}.

\subsection{Evidence for disks in fast rotators}

Kinematic sub-components with azimuthal profiles that can be fitted
with a cosine law are described as having a {\it Disk-like rotation}
(DR). This does not mean that they are actual disks. It just suggests
that velocity profiles of early-type galaxies extracted along the
best-fitting ellipse {\it resemble} the velocity maps of thin disks in
circular motion. The rate of occurrence of DRs is, however,
striking. There is no reason why this should be the case in early-type
galaxies, which in principle as a class can have a triaxial symmetry
and complex motion in different planes. As suggested above, the link
between the kinemetry assumption and the structure of fast rotating
galaxies has its origin in their internal structure.

The results of the dynamical models in Paper X reveal that fast
rotators show evidence for a kinematically distinct flattened
spheroidal component, suggesting that fast rotators are nearly oblate
and contain flattened components. In addition to these dynamically
cold components, the stellar populations of fast rotators show
evidence for different chemical
components. \citet{2006MNRAS.369..497K} find that all morphologically
flat fast rotators have Mgb line-strength distribution flatter than
the isophotes, and associate it with the rotationally supported
substructure, which features a higher metallicity and/or an increased
Mg/Fe ratio as compared to the galaxy as a whole.

These are some of the dynamical and chemical evidences for disk-like
components in fast rotators. What is the kinemetric evidence?  As
mentioned above, we find that velocity maps of fast rotators are
mostly described by a simple cosine law, as are velocity maps of thin
disks. We also find an almost one-to-one correspondence between the
projected shape of the stellar distribution and the shape of the
observed kinematic structure in fast rotating galaxies. The connection
between the shape and the kinematics is supported by an assumption
that rotation influences the shape of the object by flattening it and,
for isotropic models, the rotation speed responsible for flattening of
the shape is related to the shape of the stellar distribution as $\sim
\sqrt{\epsilon}$ \citep[][Section 4.8.2]{2008gady.book.....B}. In
order to investigate further the $\langle $\qp$\rangle - \langle
$\qk$\rangle$ correlation we constructed two-integral analytic models
of early-type galaxies.

The isotropic models we used were previously presented in Appendix B
of Paper X, to which we refer the reader for more details. The main
point of these Jeans models is that we used as templates 6 galaxies,
which represent some of the typical types from the SAURON
sample. Their light distribution was parameterised in Paper IV by the
MGE method, and was used as the basis for the intrinsic density
distributions. Observables of each Jeans model were projected at 6
different inclinations: $90\degr$, $80\degr$, $65\degr$, $50\degr$,
$35\degr$ and $25\degr$. These models are not meant to reproduce the
observed kinematics in detail, but they are self-consistent, and
under the assumption of axisymmetry and isotropy, they predict
velocity maps and offer an opportunity to study the relation between
the shape and kinematics.

On the left hand panel of Fig.~\ref{f:jeans} we show the difference
between the luminosity-weighted average values of
$\langle$\qk$\rangle$ and $\langle$\qp$\rangle$, measured by kinemetry
on the model images and velocity maps, in the same way as for the
SAURON data in Fig.~\ref{f:qkvqp} (also excluding the inner
$5\arcsec$). Different colours represent Jeans models based on
different template galaxies. Each symbol corresponds to a model at
different inclination, where the points move from top to bottom with
increasing inclination (from $25-90\degr$). It is clear that for small
inclination $\langle$\qk$\rangle$ $\approx$ $\langle$\qp$\rangle$, but
as the models are viewed closer to edge-on there is a trend of
increasing differences between the axial ratios and in some cases a
trend of larger variation along the profiles represented by larger
error bars for progressively more inclined models. Specifically, in
all but one marginal case $\langle$\qk$\rangle>\langle$\qp$\rangle$:
in these models velocity maps are `rounder' than images. The velocity
map of the Jeans model of NGC4621, whose MGE parametrisation has the
flattest Gaussian, has a flat component along the major axis, which
becomes more prominent with increasing inclination, contributing to
the radial variation and increasing \qk~with respect to \qp. The
kinematics of this component is the most 'disk-like' in our Jeans
models with tightly pinched iso-velocity contours. For comparison, the
model for NGC3377 also has a similar thin MGE component. This
component contributes to the disk-like kinematic component confined to
the central region, but since it is not as prominent in the total
light, such as the one in the NGC4621 MGE model, only traces of the
disk-like component can be seen in both photometry and kinematics.

The contrast between observed galaxies and isotropic models is
significant. The isotropic models predict $\langle$\qk$\rangle \ge
\langle$\qp$\rangle$ for $i\gtrsim 30\degr$, but the models also show
that prominent disk-like photometric features will generate pinched
iso-velocity contours and decrease $\langle$\qk$\rangle$ pushing the
galaxies towards the observed trend of, on average,
$\langle$\qk$\rangle \sim \langle$\qp$\rangle$ or even
$\langle$\qk$\rangle<\langle$\qp$\rangle$. It seems reasonable to
assume that while isotropic models can explain certain features of
fast rotators, they cannot describe them as a class of galaxies. It is
the embedded flattened components, often visible only in the
kinematics that are responsible for the observed differences between
the models and the data.

This can be also seen by comparing the difference in shape between the
isophotes and iso-$\sigma$ contours (Fig.~\ref{f:b2a0}). We repeated
the same exercise with velocity dispersion maps of our isotropic Jeans
models. Result are shown on the right panel of Fig.~\ref{f:jeans}. The
trend shown here of a larger absolute $\langle b_2/a_0 \rangle
_\sigma$ values with lower $\langle$\qp$\rangle$ is very similar to
the observed trend (hatched region). The test can also explain the
shape of the observed trend: large absolute values of $\langle b_2/a_0
\rangle _\sigma$ can be observed when the object contains a
significant flattened component and it is observed at larger
inclinations. The most affected are again NGC4621 and NGC3377
models. As before, the isotropic models are able to explain part of
the observed data, but not the details of the
distribution. Specifically, the shape of the iso-$\sigma$ contours of
slow rotators ($\langle b_2/a_0 \rangle _\sigma > 0$) is not
reproduced well by the isotropic models. Similarly, the spread in
$\langle b_2/a_0 \rangle _\sigma$ of fast rotators is also not well
reproduced. Clearly, our Jeans models are much simpler than the real
galaxies, lacking by construction multiple kinematic and especially
counter-rotating components. Comparing the kinemetric analysis of the
Jeans models and the observed objects, we find that fast rotating
galaxies are more complex than isotropic rotators, presumably
containing also flattened kinematically distinct components, which can
co- or counter-rotate on top of the non-rotating or isotropically
rotating spheroid.

The evidence for disks in fast rotators are also present in the ratio
of $h_3$ and $V/\sigma$. $h_3$ measures the asymmetric deviations from
the Gaussian LOSVD and the anti-correlation of $h_3$ with $V/\sigma$
is taken to show presence of disks. Our results confirm previous
findings that early-type galaxies on the whole have asymmetric LOSVDs,
but this applies only to fast rotators. We also find that many fast
rotators show constant and close to zero $h_3$ profiles, or, in a few
cases, show a change from negative to positive values with increasing
radius, similar to what is seen in peanut bulges and bars
\citep{2004AJ....127.3192C, 2005ApJ...626..159B}

The embedded flattened components in fast rotators are also evident
when $h_3 - V/\sigma$ diagram is compared with the results of merger
simulations \citep{1991A&A...249L...9B, 2000MNRAS.316..315B,
  2001ApJ...555L..91N, 2006MNRAS.372L..78G}. Specifically, the
updated $h_3 - V/\sigma$ diagram in Fig.~\ref{f:h3_prof} can be
compared with Fig.16 from \citet{2006MNRAS.372..839N}, who discuss the
influence of dissipational mergers in which embedded disks are formed
in merger remnants. Our figure compares rather well with a
combination of 1:1 dry and 1:3 wet mergers. This comparison suggest
that it is not possible to explain the LOSVD of early-type galaxies
with one merging track only, but that slow rotators predominantly
originate in major colissionless mergers, while fast rotators are
remnants of dissipational mergers.

The comparison of our bin-by-bin $h_4 - V/\sigma$ diagram (right-hand
panel in Fig.~\ref{f:h4}) and lower four panels of Fig.16 in
\citet{2006MNRAS.372..839N} is equally impressive, although their
simulations predict somewhat too negative values of $h_4$. Again, the
observations are largely consistent with the scenario where slow
rotators originate from dry 1:1 mergers, while fast rotators from a
combination of dry and wet 1:3 mergers.

Combining all the evidence presented in the previous section, we
suggest that fast rotators are dominated by disks (e.g. NGC3156,
NGC2685).  When their light is dominated by the bulge, their
kinematics still show strong disk components (e.g. NGC821,
NGC4660). In either case, {\it fast rotators contain flattened fast
  rotating components and this dynamical property differentiates them
  from slow rotators}. We suggest that with increasing specific
angular momentum, $\lambda_R$, the relative mass of the embedded disks
also increases and contributes more significantly to the total mass.
Among the disk-like components in fast rotators there is a range of
flattenings reflecting a diversity in possible formation paths which
create, preserve and/or thicken disks within spheroids, such as
passive fading of spirals and multiple minor dissipational
mergers. The change within the internal structure is observationally
reflected in the transition between slow and fast rotators, which
offers possible anchor points for theoretical models of galaxy
evolution.

%
%
\section{Conclusions}
\label{s:conc}

Using kinemetry we analysed two-dimensional maps of 48 early-type
galaxies observed with SAURON and velocity maps of a sub-sample
observed with OASIS. The analysed maps are: reconstructed image, mean
velocity, velocity dispersion, $h_3$ and $h_4$ Gauss-Hermite
moments. The reconstructed images and the maps of the mean velocity
were analysed along the best fit ellipses. Maps of the higher moments
of the LOSVD were analysed either along the isophotes (velocity
dispersion, $h_4$) or best-fitting ellipses from velocity maps
($h_3$). We presented the profiles of kinemetric coefficients for
velocity maps, being the dominant kinematic moment of the LOSVD and
having the highest signal-to-nose ratio.

Kinemetry and its kinemetric coefficients allow us {\it (i)} to
differentiate between slow and fast rotators from velocity maps only
and {\it (ii)} to indicate that fast rotating galaxies contain disks
with a larger range in the mass fractions to the main body. The results
of this work can be summarised as follows:

\begin{itemize}

\item Following the kinemetric analysis of the velocity maps it is
  possible to distinguish between galaxies with {\it Single} and {\it
    Multiple Components}. Components can be described as having a {\it
    Disk-like rotation}, {\it Low-level velocity}, {\it Kinematic
    misalignment}, {\it Kinematic twist} or being {\it Kinematically
    Decoupled Component}. The sorting of galaxies in these groups is
  based on kinemetric coefficients: \PAk, \qk, $k_1$ and $k_5/k_1$.

\item The majority of early-type galaxies are MC systems (69\%)

\item The total fraction of galaxies with a DR component (including
  all that have $k_5/k_1<0.02$) is 81\%. In terms of S0/E
  classification, 92\% of S0s and 74\% of Es have components with
  disk-like kinematics.

\item KDCs are found in 29\% of galaxies. These KDC are of different
  sizes, some are not resolved at SAURON resolution and appear as
  CLVs, but are clearly detectable in the OASIS observations at higher
  spatial resolution.

\item Early-type galaxies with constant \PAk ($\Delta PA <10\degr$
  outside any central component) are fast rotators. All fast-rotators
  have a DR component.

\item Most fast rotators have $\langle$\qk$\rangle \lesssim
  \langle$\qp$\rangle$. Their kinematics often show a structure
  flatter than the distribution of light. This means that images alone
  are not sufficient to recognise all fast rotators (e.g. NGC524 or
  N3379 would be missed)

\item In face-on galaxies, isophotes are coincident with iso-$\sigma$
  contours. In edge-on galaxies, however, there are differences which
  can be detected with kinemetry. Specifically, the edge-on fast
  rotators contain dynamically cold components which changes the shape
  of iso-$\sigma$ contours.

\item Slow rotating galaxies have low $V/\sigma$ and a range in $h_3$
  amplitudes. Fast rotating galaxies have large spread in $h_3$ which
  anti-correlates with $V/\sigma$. There is, however, a significant
  number of fast rotating galaxies which have $h_3$ radial profiles
  that change from strongly negative to near to zero $h_3$
  values. Similar behaviour was found in bar galaxies or in remnants
  of colissionless mergers.

\item Allowing for large uncertainties in the determination of $h_4$,
  our data show a trend where slow and fast rotators have similar
  $h_4$ values, with a weak tendency for an increased spread of $h_4$
  in fast rotators, also dependant on $V/\sigma$. These trends can be
  explained through a combination of dry and wet mergers of both equal
  and unequal in mass progenitors.

\item Dissimilarities between slow and fast rotators originate in
  their different internal structures. Slow rotators are mildly triaxial
  objects supporting a variety of orbital families. Fast rotators are
  axisymmetric spheroids with embedded flattened components of
  different mass fraction ranging from completely disk dominated
  systems to small central disks visible only in kinematics.

\end{itemize}

\vspace{+1cm}
\noindent{\bf Acknowledgements}\\

DK acknowledges the support from Queen's College Oxford and
hospitality of Centre for Astrophysics Research at the University of
Hartfordshire.  The SAURON project is made possible through grants
614.13.003, 781.74.203, 614.000.301 and 614.031.015 from NWO and
financial contributions from the Institut National des Sciences de
l'Univers, the Universit\'e Lyon~I, the Universities of Durham,
Leiden, and Oxford, the Programme National Galaxies, the British
Council, PPARC grant `Observational Astrophysics at Oxford 2002--2006'
and support from Christ Church Oxford, and the Netherlands Research
School for Astronomy NOVA.  RLD is grateful for the award of a PPARC
Senior Fellowship (PPA/Y/S/1999/00854) and postdoctoral support
through PPARC grant PPA/G/S/2000/00729. The PPARC Visitors grant
(PPA/V/S/2002/00553) to Oxford also supported this work.  GvdV
acknowledges support provided by NASA through grant NNG04GL47G and
through Hubble Fellowship grant HST-HF-01202.01-A awarded by the Space
Telescope Science Institute, which is operated by the Association of
Universities for Research in Astronomy, Inc., for NASA, under contract
NAS 5-26555. JFB acknowledges support from the Euro3D Research
Training Network, funded by the EC under contract
HPRN-CT-2002-00305. This paper is based on observations obtained at
the William Herschel Telescope, operated by the Isaac Newton Group in
the Spanish Observatorio del Roque de los Muchachos of the Instituto
de Astrof\'{\i}sica de Canarias. Based on observations obtained at the
Canada-France-Hawaii Telescope (CFHT) which is operated by the
National Research Council of Canada, the Institut National des
Sciences de l'Univers of the Centre National de la Recherche
Scientifique of France, and the University of Hawaii. This project
made use of the HyperLeda and NED databases. Part of this work is
based on data obtained from the ESO/ST-ECF Science Archive Facility.



\appendix

\section{Influence of seeing on the shape of the velocity maps}
\label{s:psf}

\begin{figure*}
        \includegraphics[width=\textwidth]{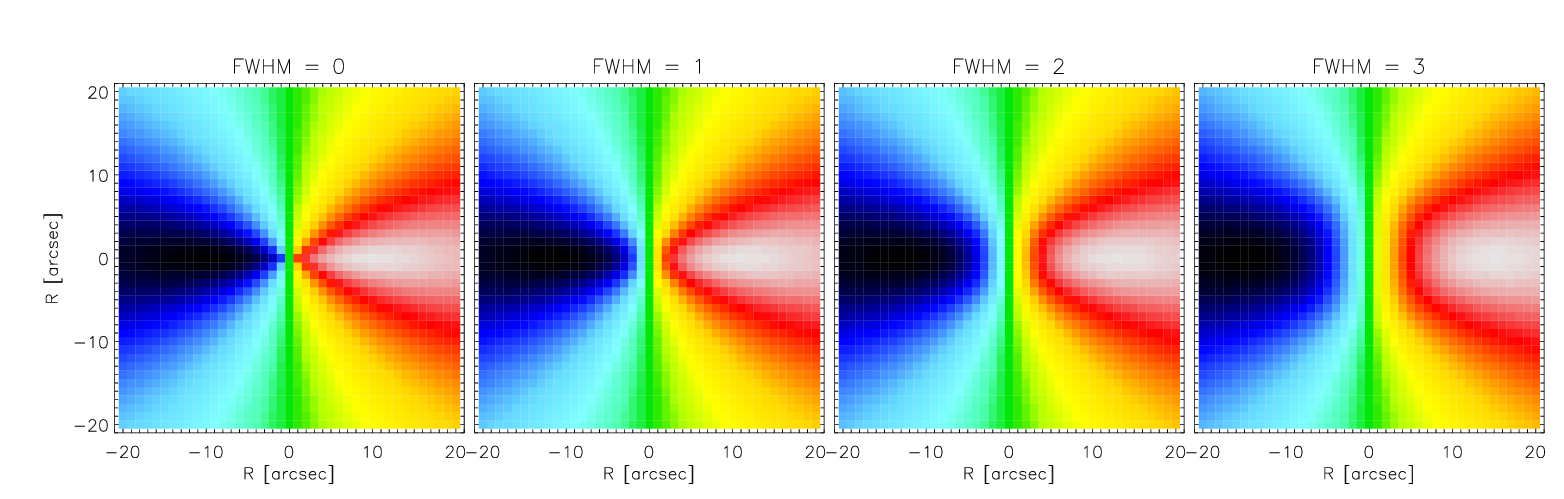}
  \caption{\label{f:maps_fwhm} An example of model velocity maps for a
    given inclination and convolved with different seeing
    kernels. {\bf From left to right:} A model velocity map at axial
    ratio of 0.5 prior to convolution with seeing. Model velocity maps
    convolved with $1\arcsec$, $2\arcsec$ and $3\arcsec$ FWHM seeing,
    respectively. Note the change in the iso-velocity contours in the
    centre with the increasing seeing.}
\end{figure*}

The effects of seeing on galaxy images in the regime of non-adaptive
optic assisted observations are now well understood
\citep{1997ApJS..109...79S}. They are specifically important in the
nuclear regions where the instrumental effects and the atmospheric
seeing redistribute photons sufficiently enough to change the
intrinsic ellipticity or position angle and mask structures like
concentrated stellar nuclei or double nuclei
\citep{1979ApJ...233...23S, 1981AJ.....86..662S,
  1983JApA....4..271D,1985ApJS...57..473L}. Specifically, luminosity
profiles are affected more by the seeing when the ellipticity is low,
while the ellipticity profiles are affected more when the ellipticity
is high \citep{1990AJ....100.1091P}. In general, effects of seeing can
be detected to quite large radii.

\begin{figure*}
        \includegraphics[width=\textwidth]{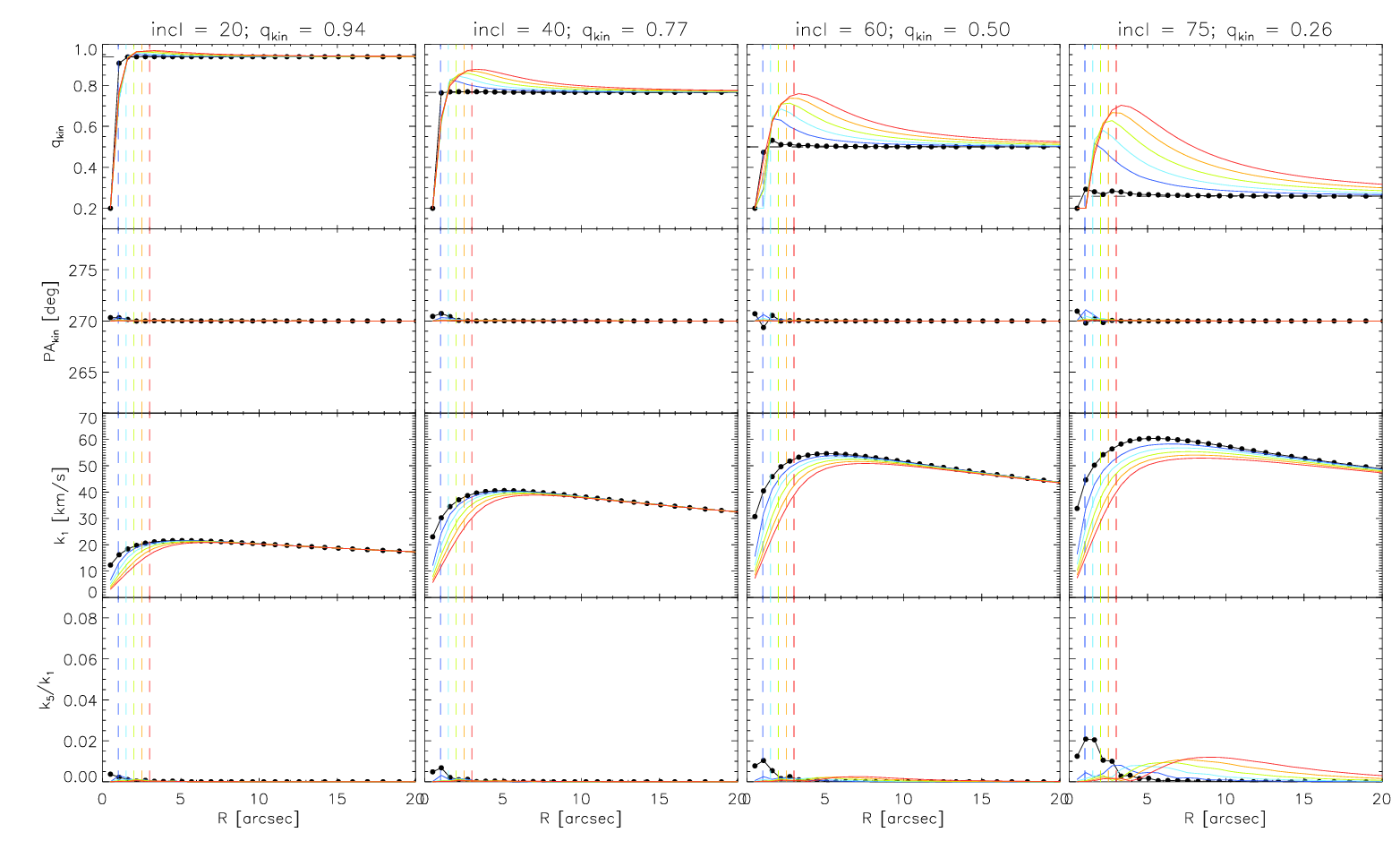}
  \caption{\label{f:prof_fwhm} Kinemetric analysis of seeing convolved
    model velocity maps.  {\bf From left to right:} Kinemetry
    coefficients from maps with axial ratios of 0.94, 0.77, 0.5 and
    0.26 degrees.  {\bf From top to bottom:} Kinemetric coefficients:
    position angle, axial ratio, $k_1$ and $k_5$/$k_1$
    harmonics. Colours correspond to values extracted from maps
    convolved to different seeing. Black: unconvolved model map. Blue:
    model map convolved with FWHM of 1\arcsec. Light blue: model map
    convolved with FWHM of 1\farcs5. Green: model map convolved with
    FWHM of 2\arcsec. Yellow: model map convolved with FWHM of
    2\farcs5. Orange: model map convolved with FWHM of
    3\arcsec. Vertical lines show the radial extent of the seeing at
    FWHM. Note that the incorrect determination of axial ratio at the
    innermost radius is an artifact of poor sampling. }
\end{figure*}

It is natural to expect the redistribution of photons due to the
seeing will also influence the two-dimensional shape (maps) of the
higher moments of the LOSVD. Specifically, for the first moment, the
mean velocity, the obvious effect of larger seeing is a less steep
rise of the velocity in the centre. Additionally, one can expect an
increased axial ratio (rounder map), or in other words, a more open
spider diagram in the central regions. Except for these qualitative
expectations there are no quantitative estimates of the influence of
seeing on the shape of velocity maps. Observations of the SAURON
sample were taken under a variety of atmospheric conditions, so
before analysing the velocity maps, we tested the effects of different
PSFs.

We constructed a number of model velocity maps using a
\citet{1990ApJ...356..359H} potential which reasonably well
approximates the density of real early-type galaxies. Each model map
was constructed from the Hernquist circular velocity profile,
projected at a certain inclination to imitate different axial ratios
seen in the sample (\qk), and weighted by the given Hernquist surface
brightness profiles. All models have the same scale length
$r_0=10\arcsec$. The inclinations ranged from 20 to 75 degrees, where
0 and 90 degrees are face on and edge on viewing angles,
respectively. This corresponds to velocity maps with axial ratios
\qk~in the range of 0.94 to 0.26. Although in the SAURON sample there
are near to edge on galaxies, their velocity maps do not resemble the
limiting case of edge on disks (\qk=0), being considerably less
flat. The above range in axial ratios were used to match the values
observed in the SAURON sample.  Maps were then convolved with a
Gaussian kernel of different full-width-half-maximum (FWHM),
accounting also for the square pixels \citep{1995MNRAS.274..602Q} in
such a way as to make the model maps similar to observations,
especially in the central regions which remain unbinned. The range of
FWHM was also matched to the measured seeing range of SAURON
observations (Paper III). Figure~\ref{f:maps_fwhm} shows an example of
model maps for axial ratio of 0.5 and seeing FWHMs of $1\arcsec$,
$2\arcsec$ and $3\arcsec$.

We ran kinemetry on the convolved model velocity maps and extracted
the same parameters as for the observed velocity maps: \PAk, \qk,
$k_{1}$ and $k_{5}/k_{1}$ harmonics. The results can be seen in
Fig.~\ref{f:prof_fwhm} and they suggest that:

\begin{itemize}

\item determination of the position angle is not influenced by the
seeing;

\item the influence of the seeing on the determined axial ratio is
  larger on maps with intrinsically smaller axial ratio. Also, the
  bigger the seeing is, the rounder (larger axial ratio) is the
  observed map.

\item the influence of the seeing on the rotation curve ($k_1$ term)
  is larger on maps with intrinsically smaller axial ratios, where the
  bigger is the seeing, the lower is the maximum velocity. The
  position of maximum velocity is also pushed towards larger radii.

\item the influence of the seeing on the higher terms ($k_{5}/k_{1}$)
  is negligible in most cases, except for intrinsically very flat
  maps, but even there the signal is of the order of 1\% or less and
  is not detectable with the current accuracy.

\end{itemize}

As expected, the seeing influences the axial ratio of velocity maps
(or, in other words, the opening of the iso-velocity contours), and
the steepness of the rotation curve, both in the maximum values and
the positions of these maxima.

The largest effect is for maps with small axial ratio (large opening
angles), which can be understood as the consequence of the
convolution, similar as in the case of photometry and high
ellipticity. In terms of changing of the intrinsic values, it is the
axial ratio that is the most affected by seeing. At an intrinsic
flattening of 0.77, a bad seeing of $2\farcs5$, will increase the
flattening by 0.1. This effect rises to surprising 0.5 difference in
measured flattening for an intrinsic flattening of 0.26. This effect
is milder for the maximum rotation velocity, where only for velocity
maps with the smallest axial ratios and strongest seeings the
difference between the intrinsic and seeing convolved values exceeds
10 km/s. Another effect of seeing on the velocity maps is the change
of position of the maximum axial ratio and the maximum rotational
velocity with increasing seeing. In both cases the trend is clear:
larger seeing FWHM causes displacement of the maximum (\qk~or $k_1$)
towards larger radii.

In summary, seeing can strongly influence the appearance of the
velocity maps, both in the shape, value and extent of the
features. This is important for interpreting the structures on the
maps and, especially, when looking for kinematic subcomponents.

\section{Kinemetric profiles}
\label{s:profiles}

In this section we present the results of kinemetric analysis of
SAURON velocity maps.

\begin{figure*}
        \includegraphics[width=0.98\textwidth,bb=30 45 780 500]{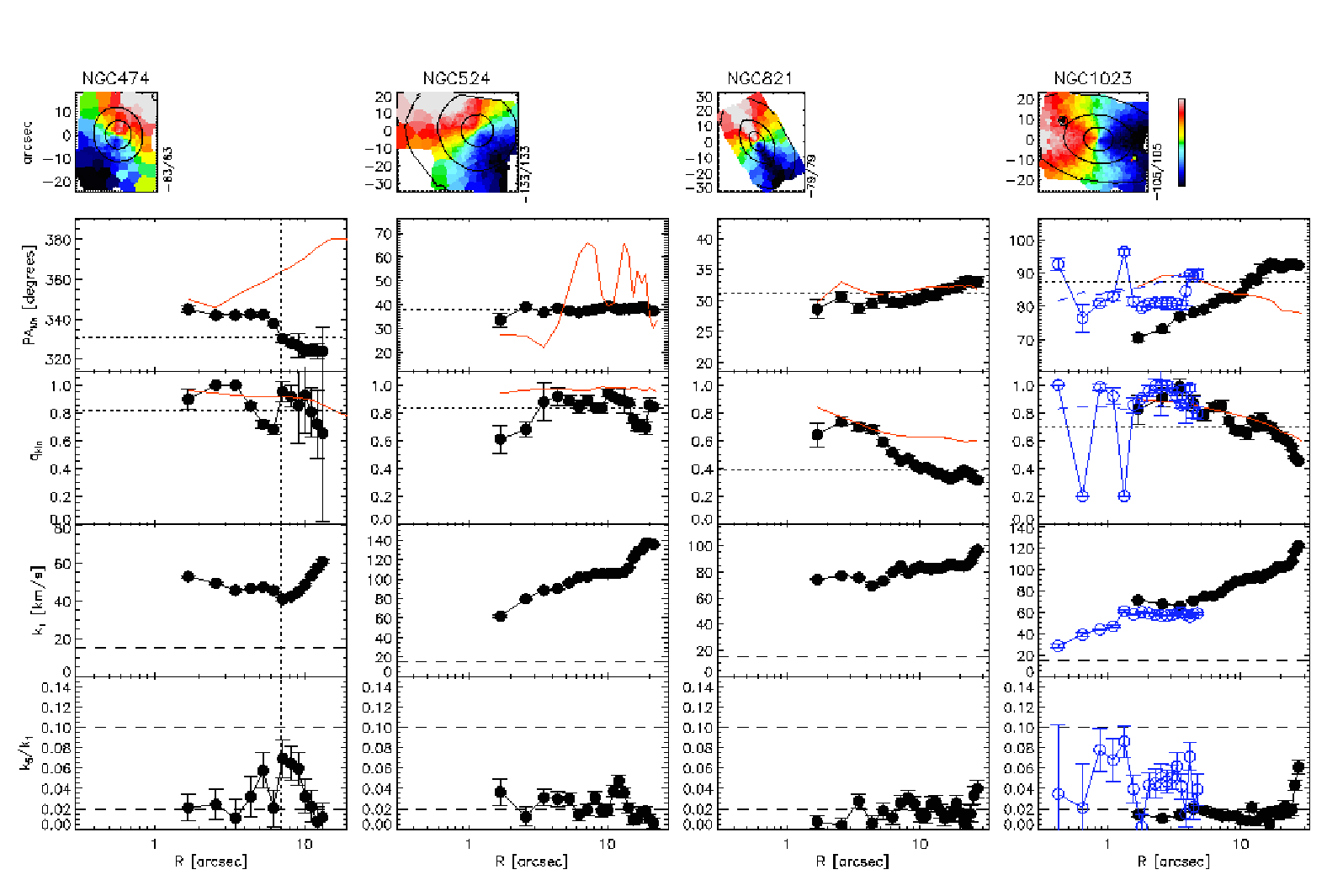}
        \includegraphics[width=0.98\textwidth,bb=30 45 780 540]{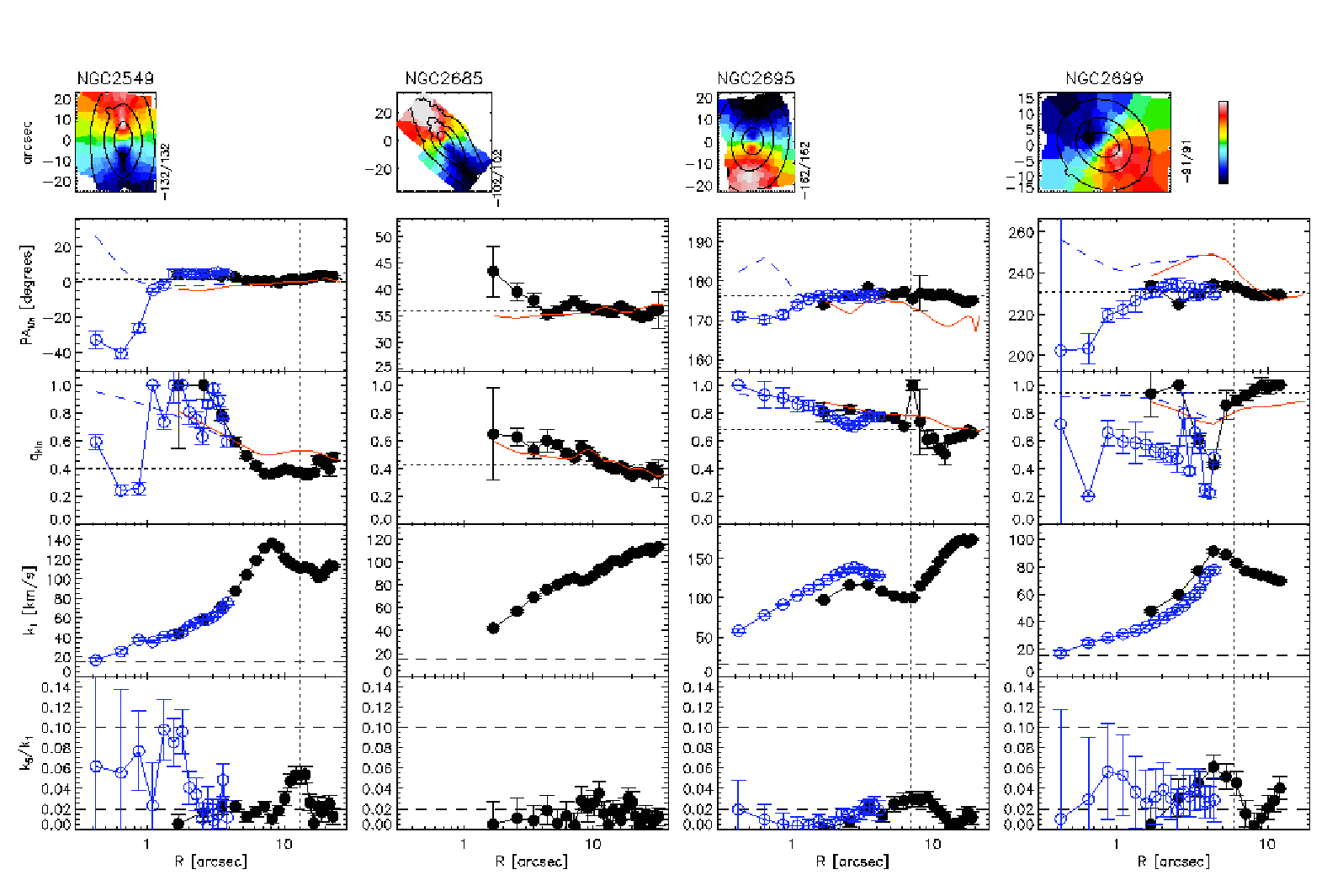}
  \caption{\label{f:vel_prof} Kinemetric profiles of 48 early-type
    galaxies. The panels show (from top to bottom) SAURON stellar mean
    velocity map, and extracted kinemetric properties: position angle,
    axial ratio, $k_1$ and $k_5$/$k_1$ coefficients. Solid black
    symbols are SAURON data, open blue circles are OASIS data. Solid
    light-blue and green symbols are results of kinemetry on circles
    for OASIS and SAURON data, respectively. Red line shows
    photometric \PAp~and \qp~from the SAURON reconstructed
    images. Dashed-blue line shows photometric \PAp~and $1-\epsilon$
    from the OASIS reconstructed images. Dashed horizontal lines on
    $k_1$ and $k_5/k_1$ panels show the limiting cases: $k_1=15 \kms$
    limit of detectable rotation, $k_5/k_1=0.02$ limit of detectable
    deviation from the assumed cosine function in kinemetry,
    $k_5/k_1=0.1$ limit for doing kinemetry on circles. These limits
    are data dependant (for more details see
    Section~\ref{s:res}). Dashed lines on \PAk~and \qk~panels show
    luminosity weighted average values of the radial profiles. In the
    case of global $\langle$\qk$\rangle$ innermost $5\arcsec$ are
    excluded due to the seeing effects. }
    
\end{figure*}
\addtocounter{figure}{-1} 
\begin{figure*}
        \includegraphics[width=\textwidth,bb=30 45 780 510]{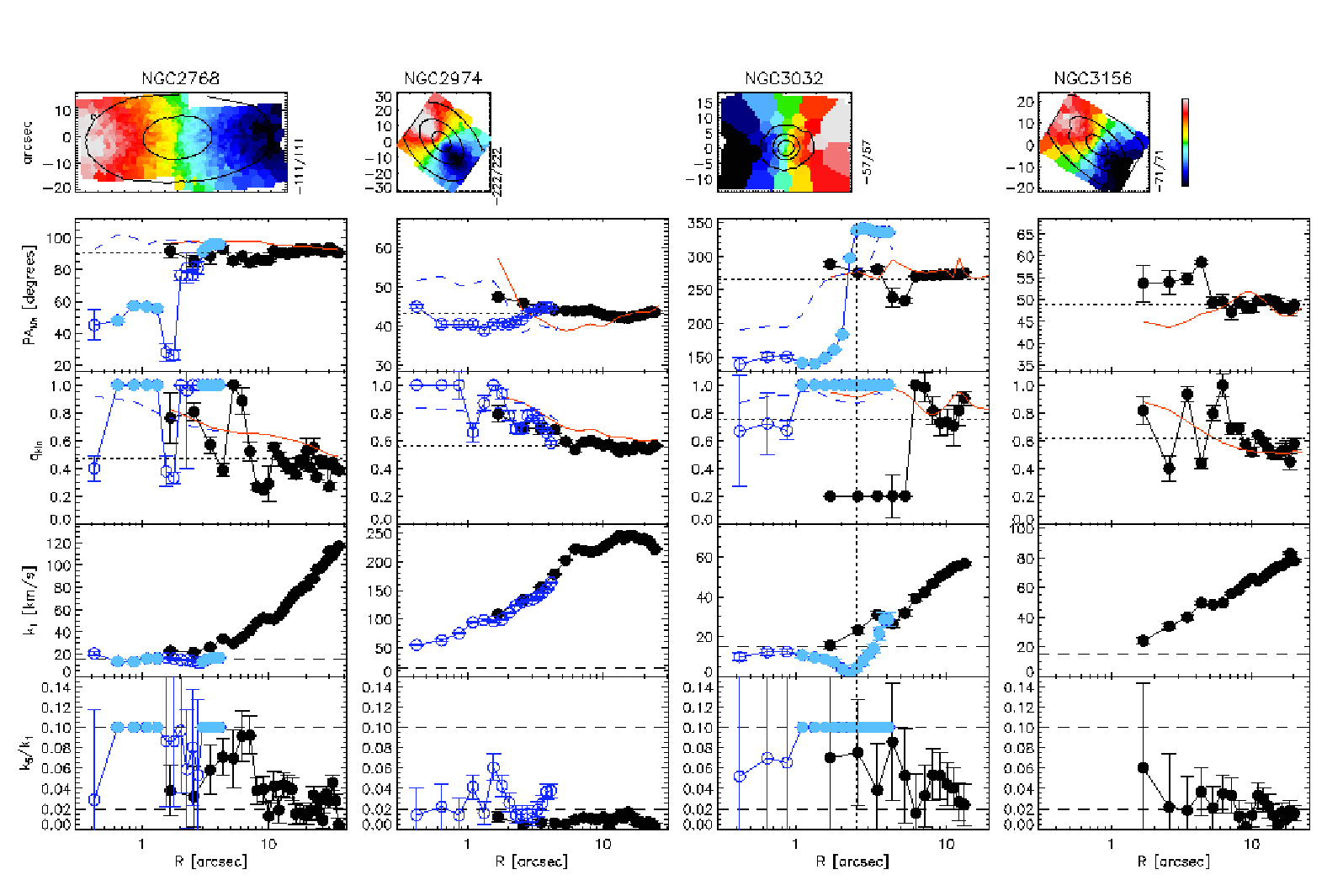}
        \includegraphics[width=\textwidth,bb=35 45 780 540]{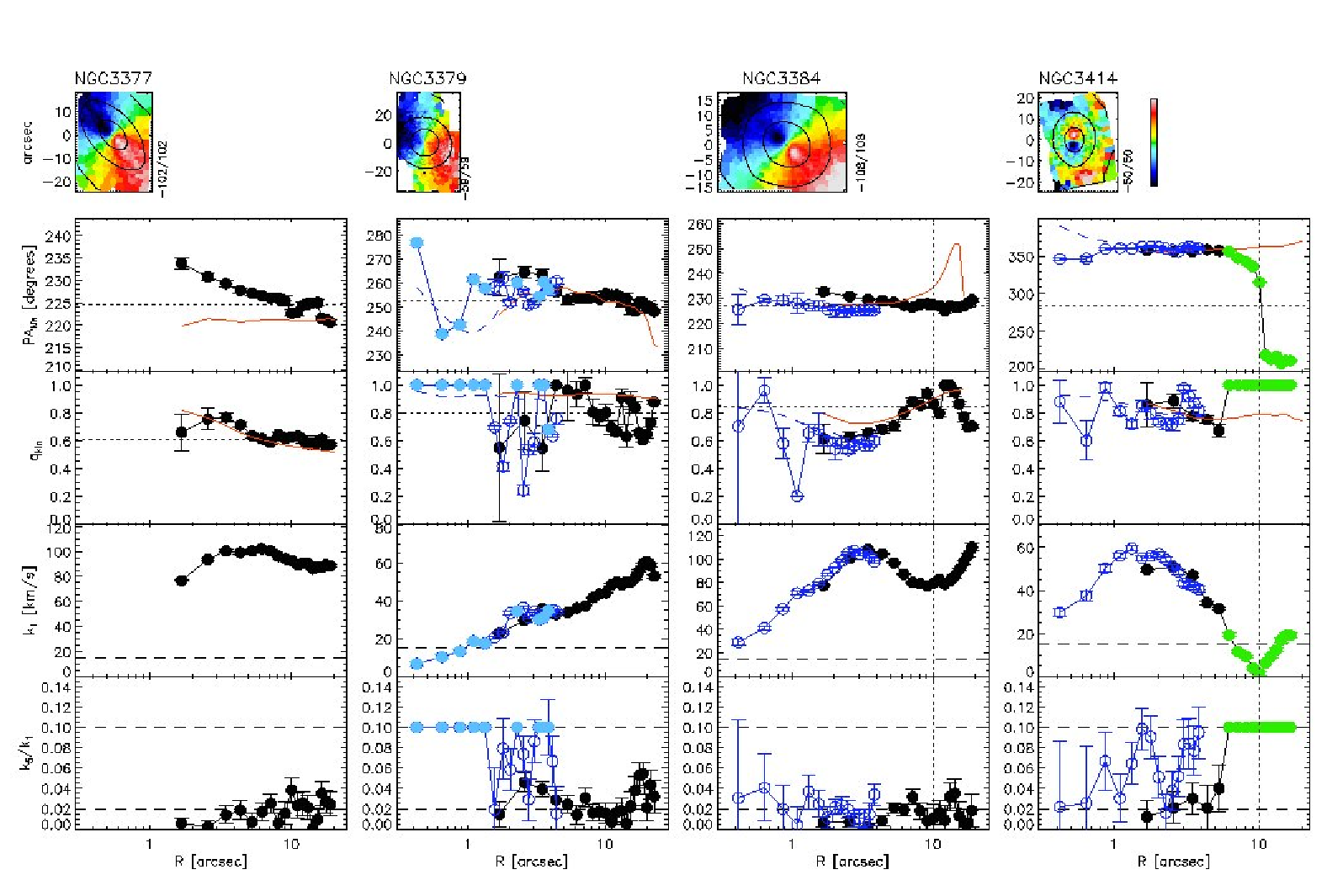}
 \caption{\label{f:prof}}
\end{figure*}
\addtocounter{figure}{-1}                                                       
\begin{figure*}
        \includegraphics[width=\textwidth,bb=30 45 780 510]{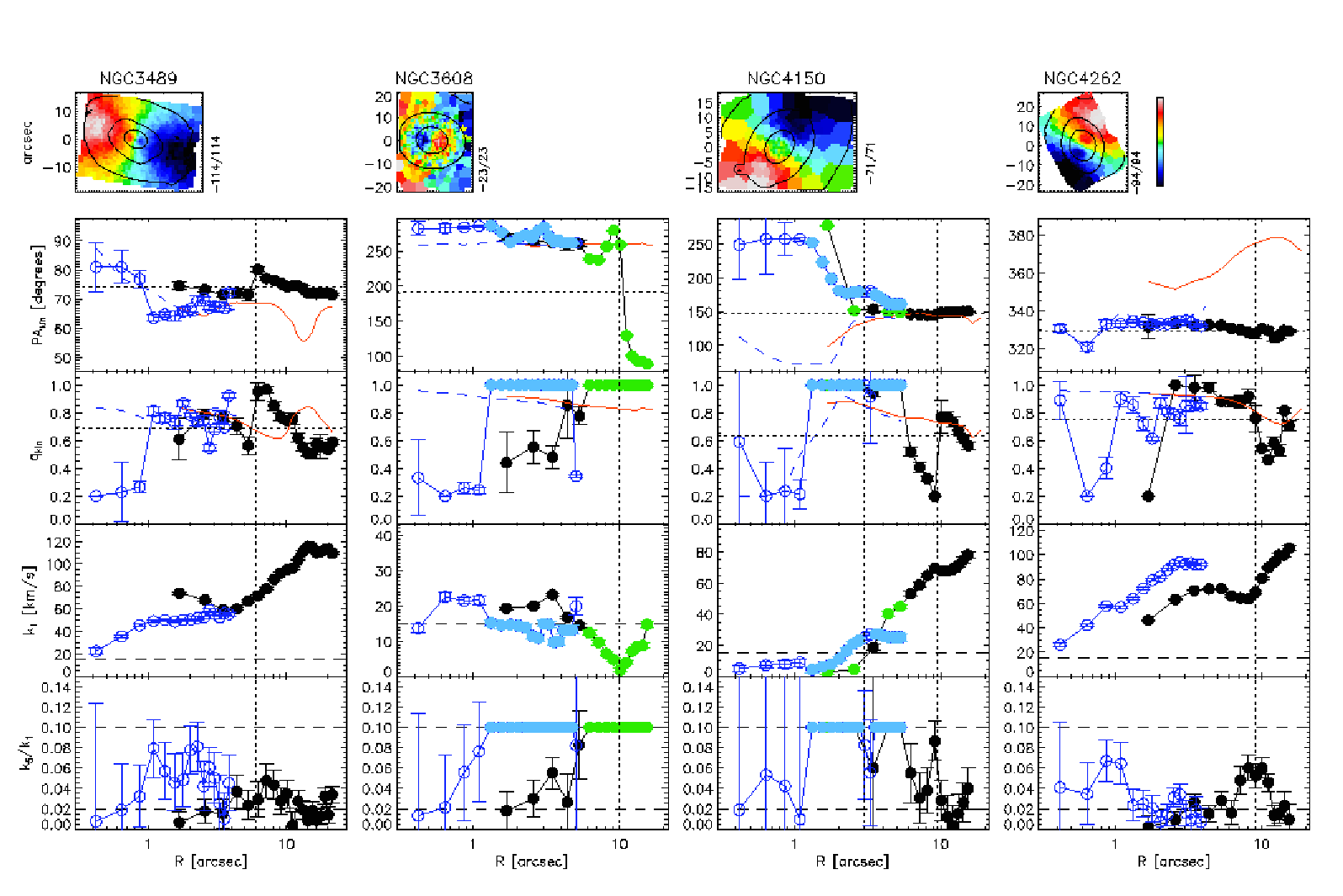}
        \includegraphics[width=\textwidth,bb=35 45 780 540]{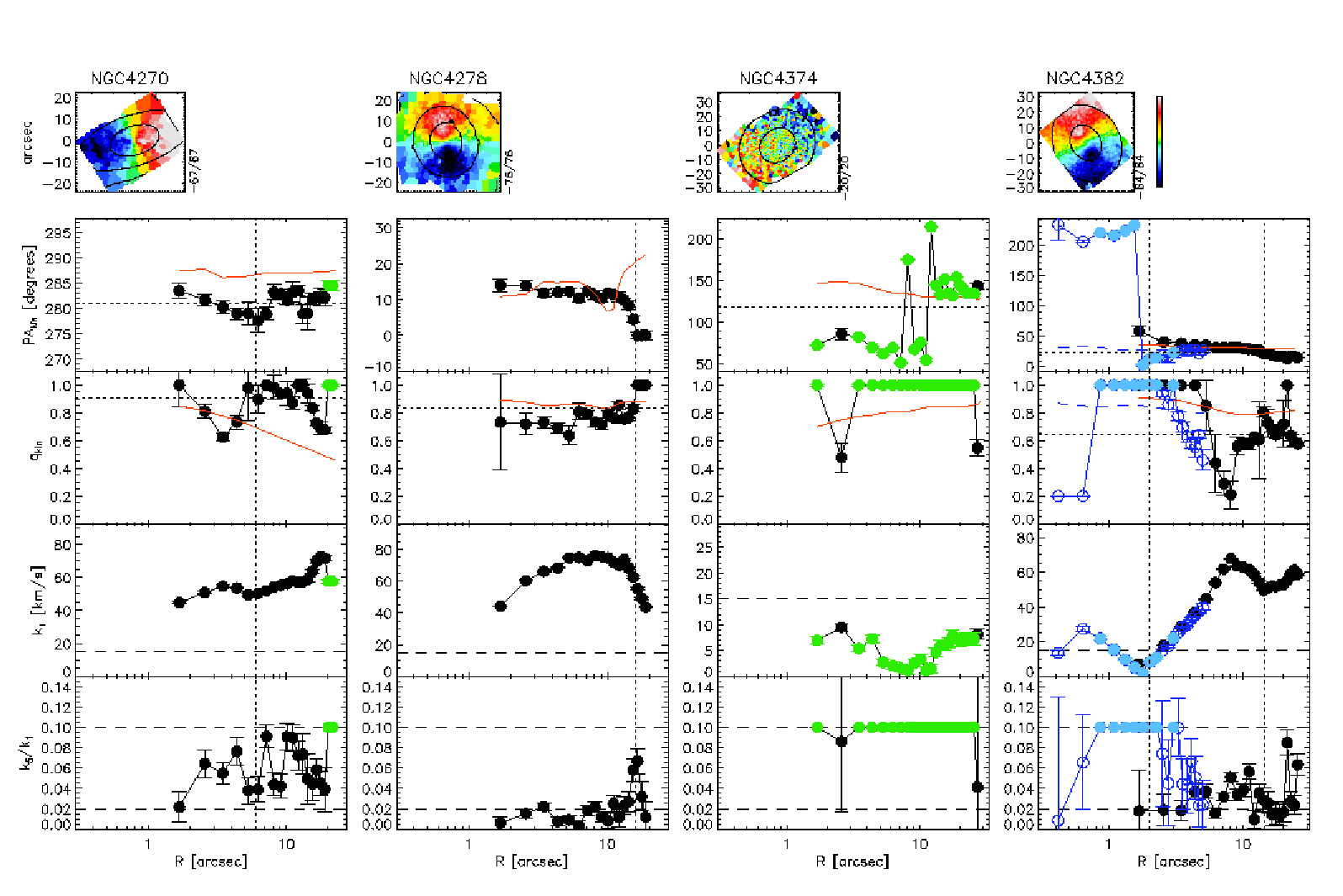}
\caption{\label{f:prof} }
\end{figure*}
\addtocounter{figure}{-1}                                                       
\begin{figure*}
        \includegraphics[width=\textwidth,bb=30 45 780 510]{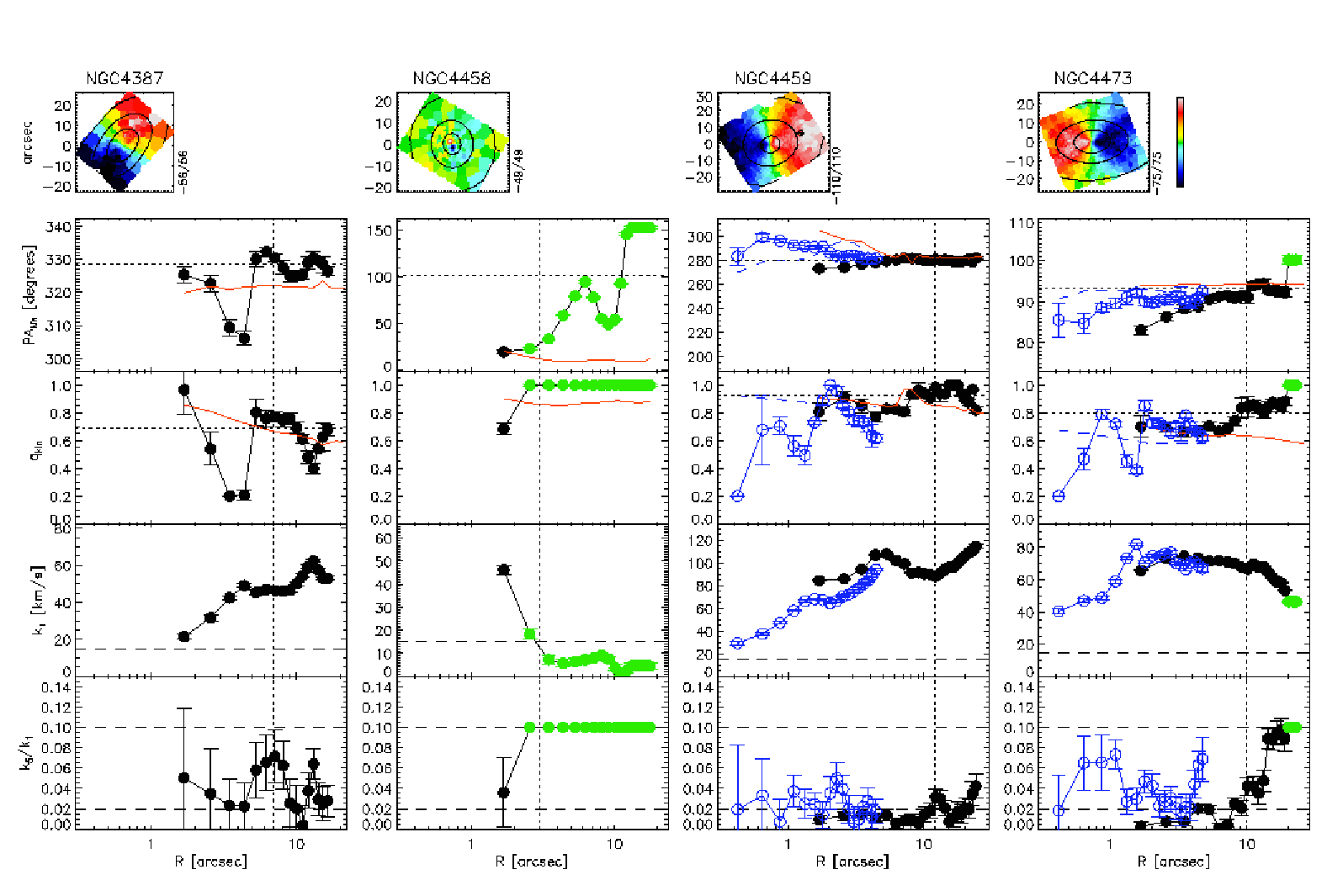}
        \includegraphics[width=\textwidth,bb=35 45 780 540]{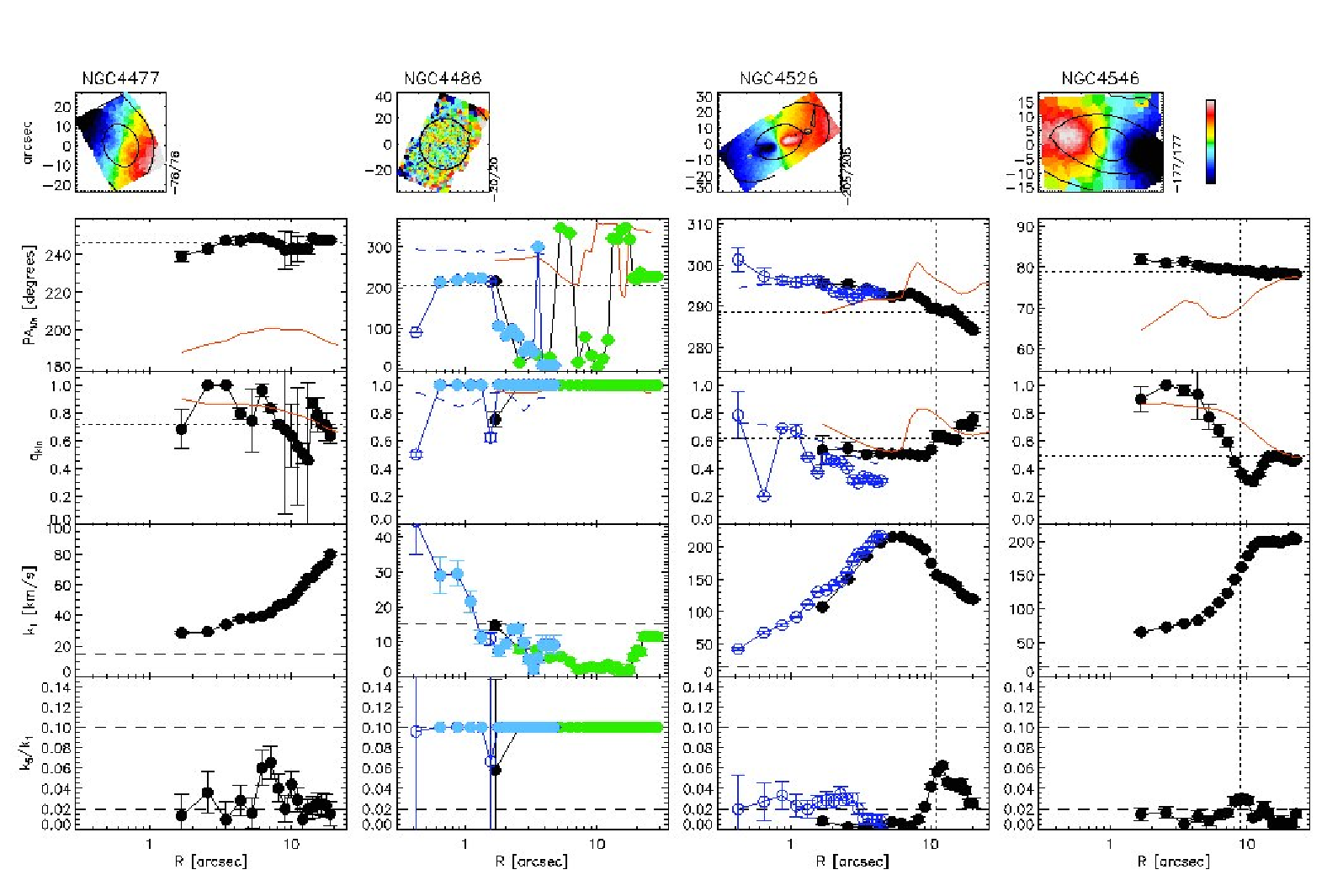}
\caption{\label{f:prof}}
\end{figure*}
\addtocounter{figure}{-1}                                                       
\begin{figure*}
        \includegraphics[width=\textwidth,bb=30 45 780 510]{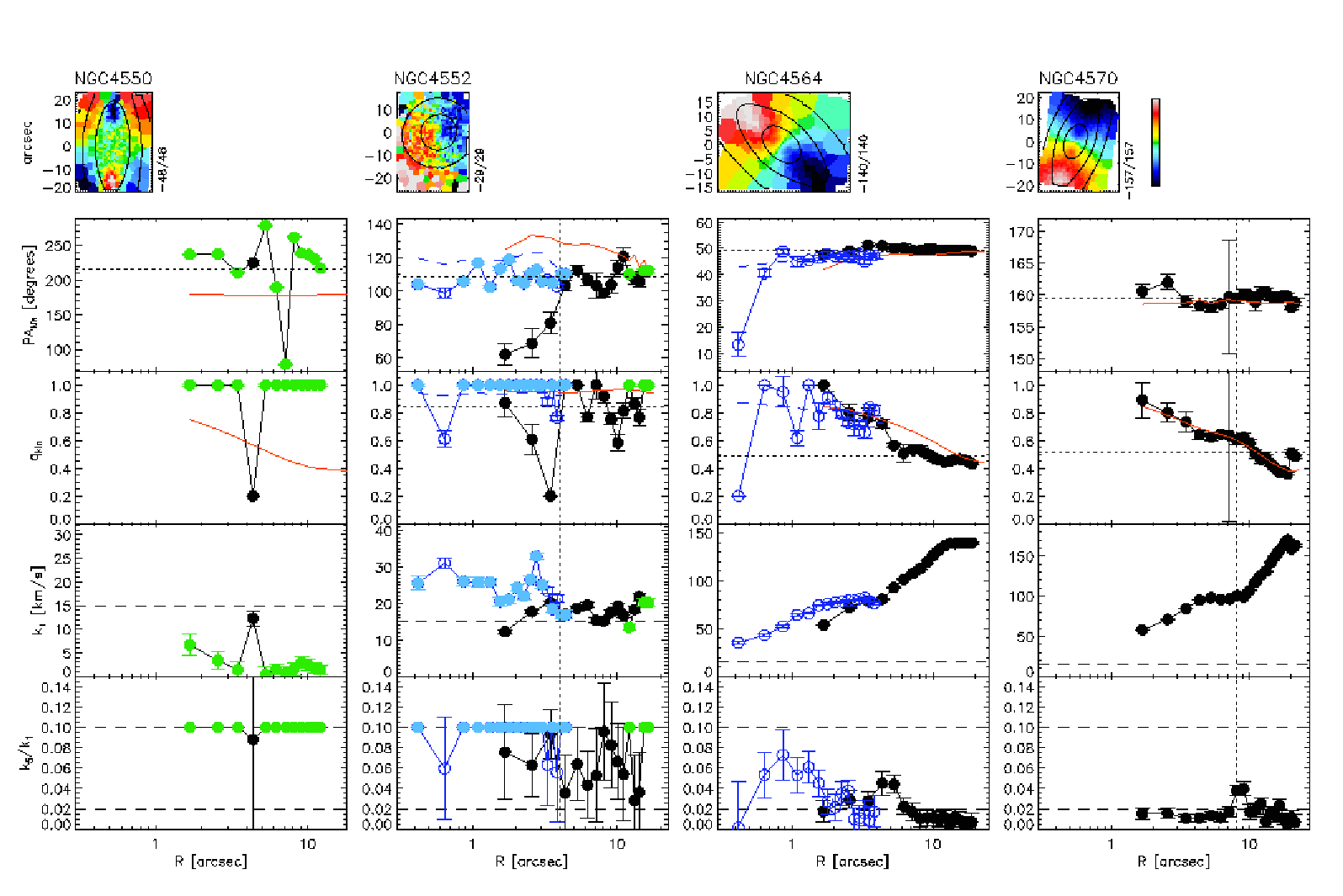}
        \includegraphics[width=\textwidth,bb=35 45 780 540]{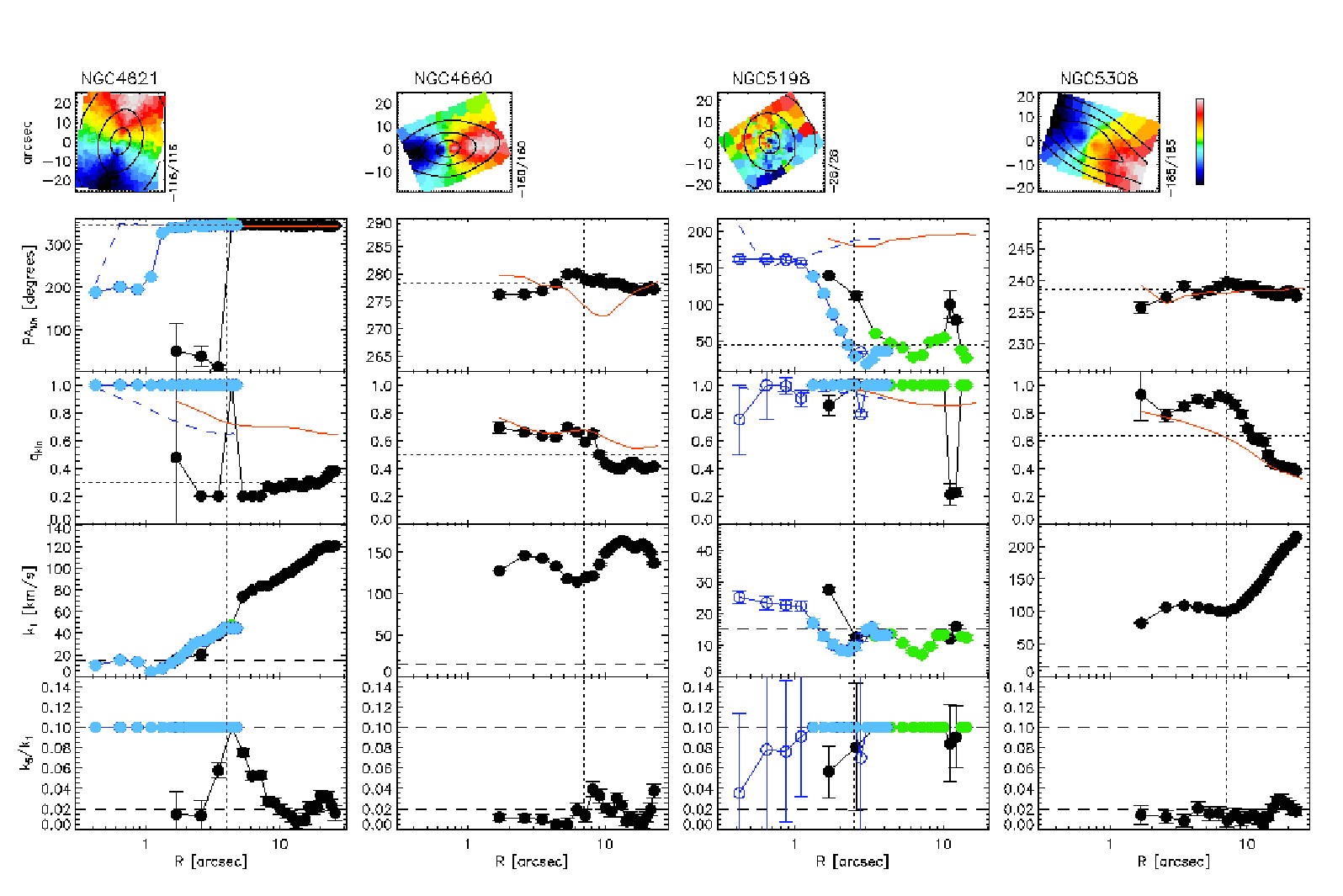}
\caption{\label{f:prof}}
\end{figure*}
\addtocounter{figure}{-1}                                                       
\begin{figure*}
        \includegraphics[width=\textwidth,bb=30 45 780 510]{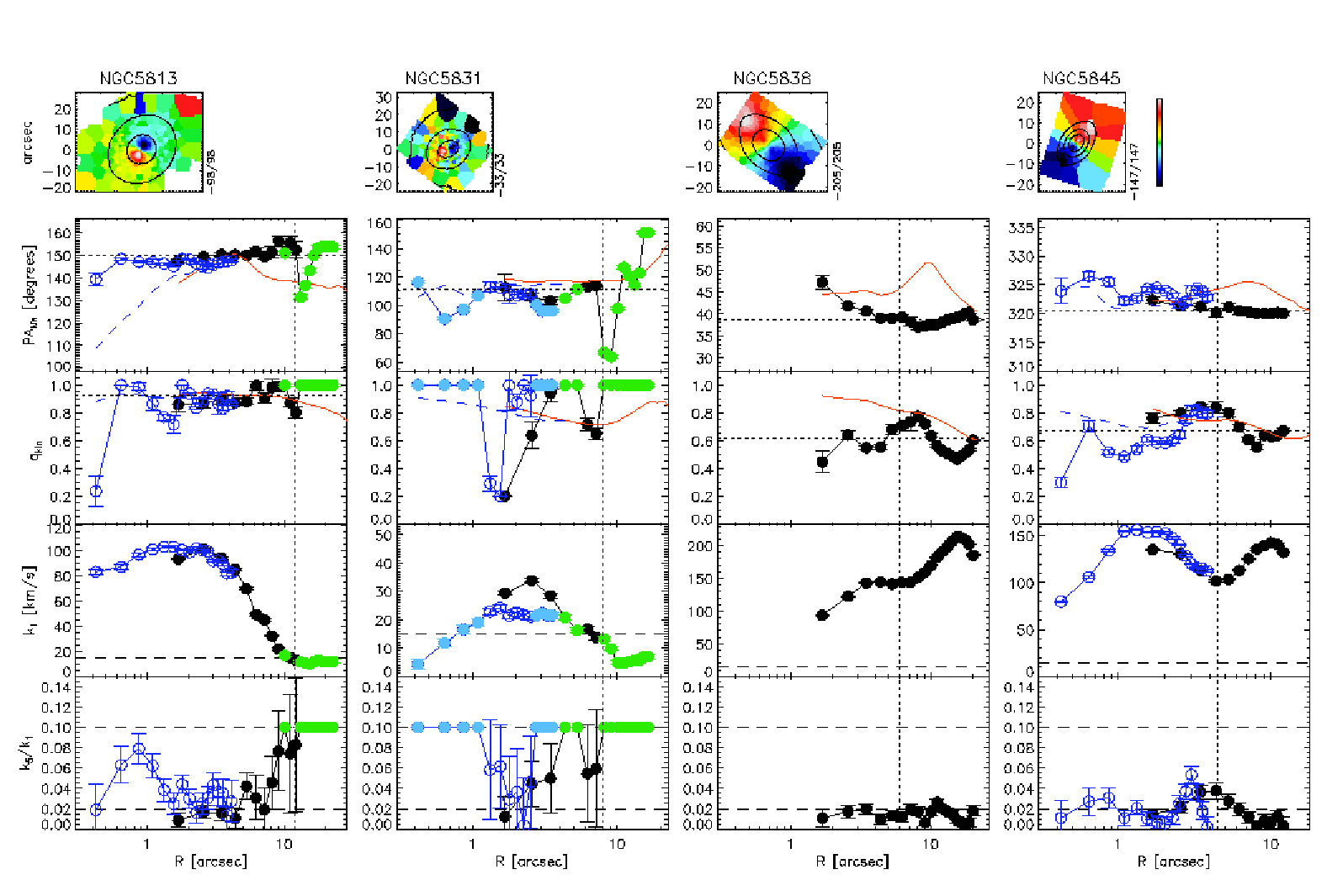}
        \includegraphics[width=\textwidth,bb=35 45 780 540]{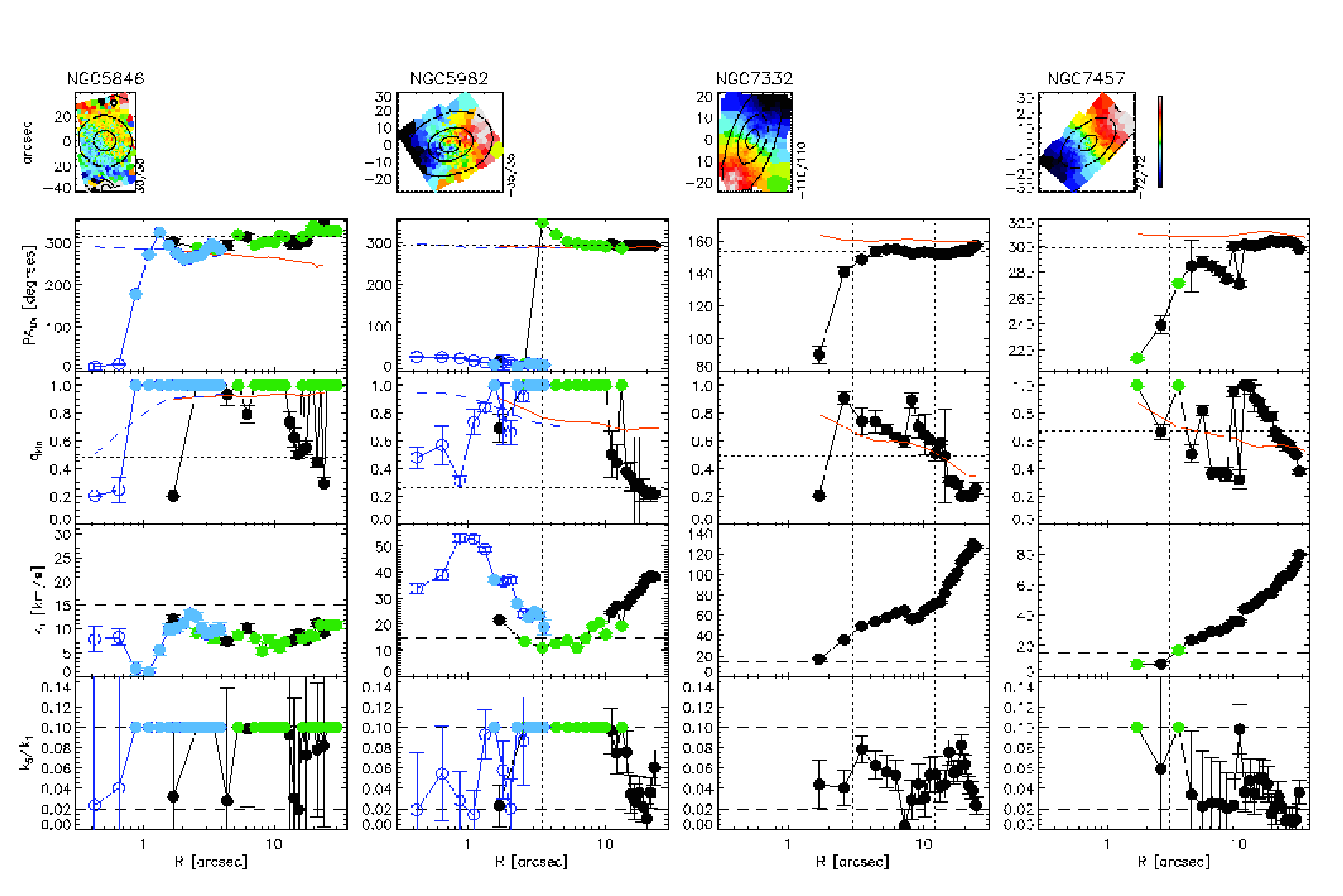}
\caption{\label{f:prof}}
\end{figure*}

\label{lastpage}

\end{document}